\documentclass[12pt]{article}
\usepackage{graphicx}
\usepackage{epsfig,cite}

\catcode`@=11

\begin{document}



\newcommand{\cp}{{\mathcal CP}}
\newcommand{\tb}{\tan\beta}
\newcommand{\MZ}{M_Z}
\newcommand{\MA}{M_A}
\def\citere#1{\mbox{Ref.~\cite{#1}}}
\def\citeres#1{\mbox{Refs.~\cite{#1}}}
\newcommand{\Stop}{\tilde t}
\newcommand{\Sbot}{\tilde b}
\newcommand{\StopL}{\tilde{t}_L}
\newcommand{\StopR}{\tilde{t}_R}
\newcommand{\SbotL}{\tilde{b}_L}
\newcommand{\SbotR}{\tilde{b}_R}
\newcommand{\ML}{\left( \begin{array}{cc}}
\newcommand{\MR}{\end{array} \right)}
\newcommand{\Mh}{m_h}
\newcommand{\mh}{m_h}
\newcommand{\MH}{m_H}
\newcommand{\mhmax}{\mh^{\rm max}}
\newcommand{\gev}{\,\, \mathrm{GeV}}
\newcommand{\tev}{\,\, \mathrm{TeV}}
\newcommand{\MstL}{M_{\tilde{t}_L}}
\newcommand{\MstR}{M_{\tilde{t}_R}}
\newcommand{\MsbL}{M_{\tilde{b}_L}}
\newcommand{\MsbR}{M_{\tilde{b}_R}}
\newcommand{\msusy}{M_{\rm SUSY}}
\newcommand{\At}{A_t}
\newcommand{\Ab}{A_b}
\newcommand{\Xt}{X_t}
\newcommand{\Xb}{X_b}
\newcommand{\mt}{m_t}
\newcommand{\mb}{m_b}
\newcommand{\mgl}{m_{\tilde{g}}}
\newcommand{\edz}{\frac{1}{2}}
\newcommand{\CZb}{\cos 2\beta}
\newcommand{\CTb}{\cot \beta\hspace{1mm}}
\newcommand{\sw}{s_\mathrm{w}}
\newcommand{\cw}{c_\mathrm{w}}
\newcommand{\drbar}{$\overline{\rm{DR}}$}
\newcommand{\BE}{\begin{equation}}
\newcommand{\EE}{\end{equation}}
\newcommand{\BEA}{\begin{eqnarray}}
\newcommand{\EEA}{\end{eqnarray}}
\newcommand{\non}{\nonumber}
\def\aeff{\al_{\rm eff}}
\newcommand{\al}{\alpha}
\newcommand{\ga}{\gamma}
\newcommand{\Ga}{\Gamma}
\newcommand{\SM}{\mathrm {SM}}
\newcommand{\MSSM}{\mathrm {MSSM}}
\def\order#1{${\cal O}(#1)$}
\newcommand{\KL}{\left(}
\newcommand{\KR}{\right)}

\def\mpar#1{\marginpar{\tiny #1}}
\def\mua{\marginpar[\boldmath\hfil$\uparrow$]%
                   {\boldmath$\uparrow$\hfil}%
                    \typeout{marginpar: $\uparrow$}\ignorespaces}
\def\mda{\marginpar[\boldmath\hfil$\downarrow$]%
                   {\boldmath$\downarrow$\hfil}%
                    \typeout{marginpar: $\downarrow$}\ignorespaces}
\def\mla{\marginpar[\boldmath\hfil$\rightarrow$]%
                   {\boldmath$\leftarrow $\hfil}%
                    \typeout{marginpar: $\leftrightarrow$}\ignorespaces}
\def\Mua{\marginpar[\boldmath\hfil$\Uparrow$]%
                   {\boldmath$\Uparrow$\hfil}%
                    \typeout{marginpar: $\Uparrow$}\ignorespaces}
\def\Mda{\marginpar[\boldmath\hfil$\Downarrow$]%
                   {\boldmath$\Downarrow$\hfil}%
                    \typeout{marginpar: $\Downarrow$}\ignorespaces}
\def\Mla{\marginpar[\boldmath\hfil$\Rightarrow$]%
                   {\boldmath$\Leftarrow $\hfil}%
                    \typeout{marginpar: $\Leftrightarrow$}\ignorespaces}


\thispagestyle{empty}
\setcounter{page}{0}
\def\thefootnote{\fnsymbol{footnote}}

\begin{flushright}
CP3--06--06\\
DCPT/06/106\\
IPPP/06/53\\
MPP--2006--93\\
hep-ph/0607308\\
\end{flushright}

\mbox{}\vspace{0em}

\begin{center}

{\large\sc {\bf SM and MSSM Higgs Boson Production Cross Sections}}

\vspace*{0.3cm}

{\large\sc {\bf at the Tevatron and the LHC}}
\footnote{
Contribution to the {\em Tev4LHC workshop}, 2005--2006.
}

\vspace{1.5cm}

{\sc T.~Hahn$^{\,1}$%
\footnote{
email: hahn@feynarts.de
}%
, S.~Heinemeyer$^{\,2}$%
\footnote{
email: Sven.Heinemeyer@cern.ch
}%
, F.~Maltoni$^{\,3}$%
\footnote{
email: maltoni@fyma.ucl.ac.be
}%
, G.~Weiglein$^{\,4}$%
\footnote{
email: Georg.Weiglein@durham.ac.uk
}%
, S.~Willenbrock$^{\,5}$%
\footnote{
email: willen@uiuc.edu
}%
}

\vspace*{0.8cm}

$^1$ Max-Planck-Institut f\"ur Physik, F\"ohringer Ring 6,
D-80805 M\"unchen, Germany 

\vspace*{0.3cm}

$^2$ Instituto de Fisica de Cantabria (CSIC--UC), 
Santander,  Spain 

\vspace*{0.3cm}

$^3$ Centre for Particle Physics and Phenomenology (CP3) \\
Universit\'{e} Catholique de Louvain\\
Chemin du Cyclotron 2
B-1348 Louvain-la-Neuve, Belgium

\vspace*{0.3cm}

$^4$ IPPP, University of Durham, Durham DH1~3LE, UK

\vspace*{0.3cm}

$^5$ Department of Physics, University of Illinois at Urbana-Champaign \\
1110 West Green Street, Urbana, IL\ \ 61801

\end{center}

\vspace*{0.8cm}

\begin{abstract}
We present results for the SM and MSSM Higgs-boson production cross
sections at  
the Tevatron and the LHC. The SM cross sections are a compilation of
the state-of-the-art theoretical predictions. The MSSM cross
sections are obtained from the SM ones by means of an effective
coupling approximation, as implemented in {\tt FeynHiggs}.  Numerical
results have been obtained in four benchmark scenarios for two values
of $\tan\beta$, $\tan\beta = 5, 40$.

\end{abstract}

\def\thefootnote{\arabic{footnote}}
\setcounter{footnote}{0}

\newpage


\section{Introduction}

Deciphering the mechanism of electroweak symmetry breaking (EWSB) is
one of the main quests of the high energy physics community.
Electroweak precision data in combination with the direct top-quark 
mass measurement at the Tevatron have
strongly constrained the range of possible scenarios and hinted to the
existence of a light scalar particle~\cite{LEPEWWG}. 
Both in the standard model (SM) and in its minimal 
supersymmetric extensions (MSSM), the $W$ and $Z$
bosons and fermions acquire masses by coupling to the vacuum
expectation value(s) of scalar SU(2) doublet(s), via the so-called
Higgs mechanism. The common prediction of such models is the
existence of at least one scalar state, the Higgs boson.  Within the
SM, LEP has put a lower bound on the Higgs mass, $\mh >
114.4$ GeV~\cite{LEPHiggsSM}, and has contributed to the
indirect evidence that the Higgs boson should be relatively light 
with a 95\% probability for its mass to be below 166 GeV~\cite{LEPEWWG}.  
In the MSSM the experimental lower bound for the mass of the lightest
state is somewhat 
weaker, while the theory predicts an upper bound
of about 135~GeV~\cite{mhiggslong,mhiggsAEC,Allanach:2004rh}.

If the Higgs sector is realized as implemented in the SM or the MSSM,
at least one Higgs boson
should be discovered at the Tevatron and/or at the LHC. Depending on the mass,
there are various channels available where Higgs searches can be performed.
The power of each signature depends on the production
cross section, $\sigma$, and the Higgs branching ratio into final
state particles, such as leptons or $b$-jets, 
the total yield of events being proportional to $\sigma \cdot$ BR. 
In some golden channels, such as $gg\to h \to Z^{(*)}Z \to 4\mu$,  a
discovery will be straightforward and mostly independent from our
ability to predict signal and/or backgrounds. On the other hand,
for coupling measurements or for searches in more difficult channels, 
such as $t\bar t h\to t \bar t b \bar b$ associated production, 
precise predictions for both signal and backgrounds are mandatory. 
Within the MSSM such precise predictions for signal and backgrounds
are necessary in order to relate the experimental results to the
underlying SUSY parameters.

The aim of this note is to collect up-to-date predictions for 
the most relevant signal cross sections, for both the SM and the MSSM. 
In Section~\ref{sec:SM} we collect the results 
of state-of-the-art calculations for the SM cross sections as a function
of the Higgs mass.  In Section~\ref{sec:MSSM} we present the MSSM 
cross sections for the neutral Higgs-bosons in four benchmark scenarios.
These results are obtained by rescaling the SM cross sections presented
in the previous sections, using an effective coupling approximation.


\section{SM Higgs production cross sections}
\label{sec:SM}

\begin{figure}[t]
\begin{center}
\includegraphics[angle=-90,width=16cm]{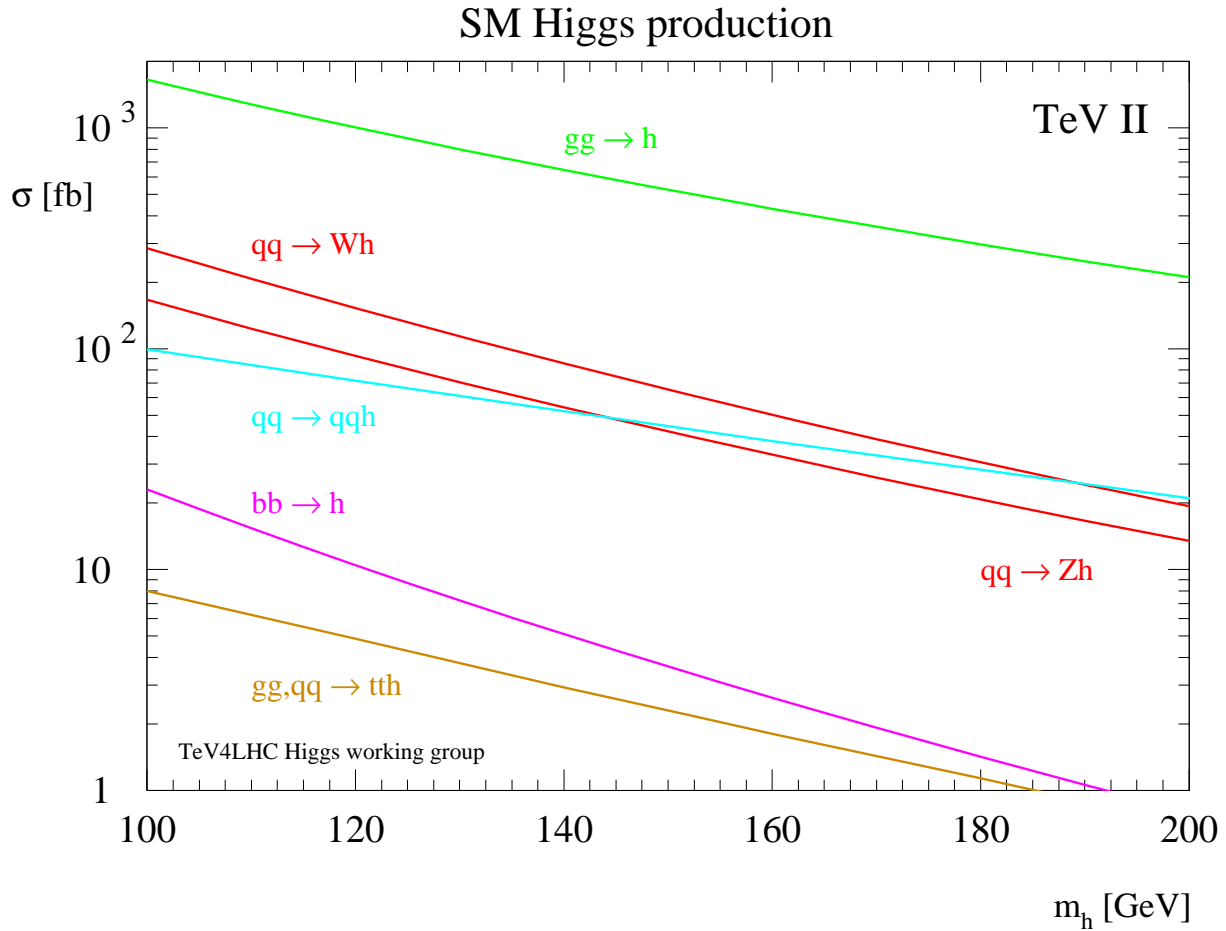}
\end{center}
\caption{Higgs-boson production cross sections (fb) at the Tevatron
($\sqrt{s}=1.96 \tev$) for the most relevant production mechanisms as a
function of the Higgs-boson mass. Results for $gg \to h$, 
$q\bar q\to Vh$, $b\bar b \to h$ are at NNLO in the QCD expansion.
Weak boson fusion ($qq \to qq h$) and $t\bar t $ associated
production are at NLO accuracy.}
\label{fig:tev}
\vspace{1em}
\end{figure}

\begin{figure}[t]
\begin{center}
\includegraphics[angle=-90,width=16cm]{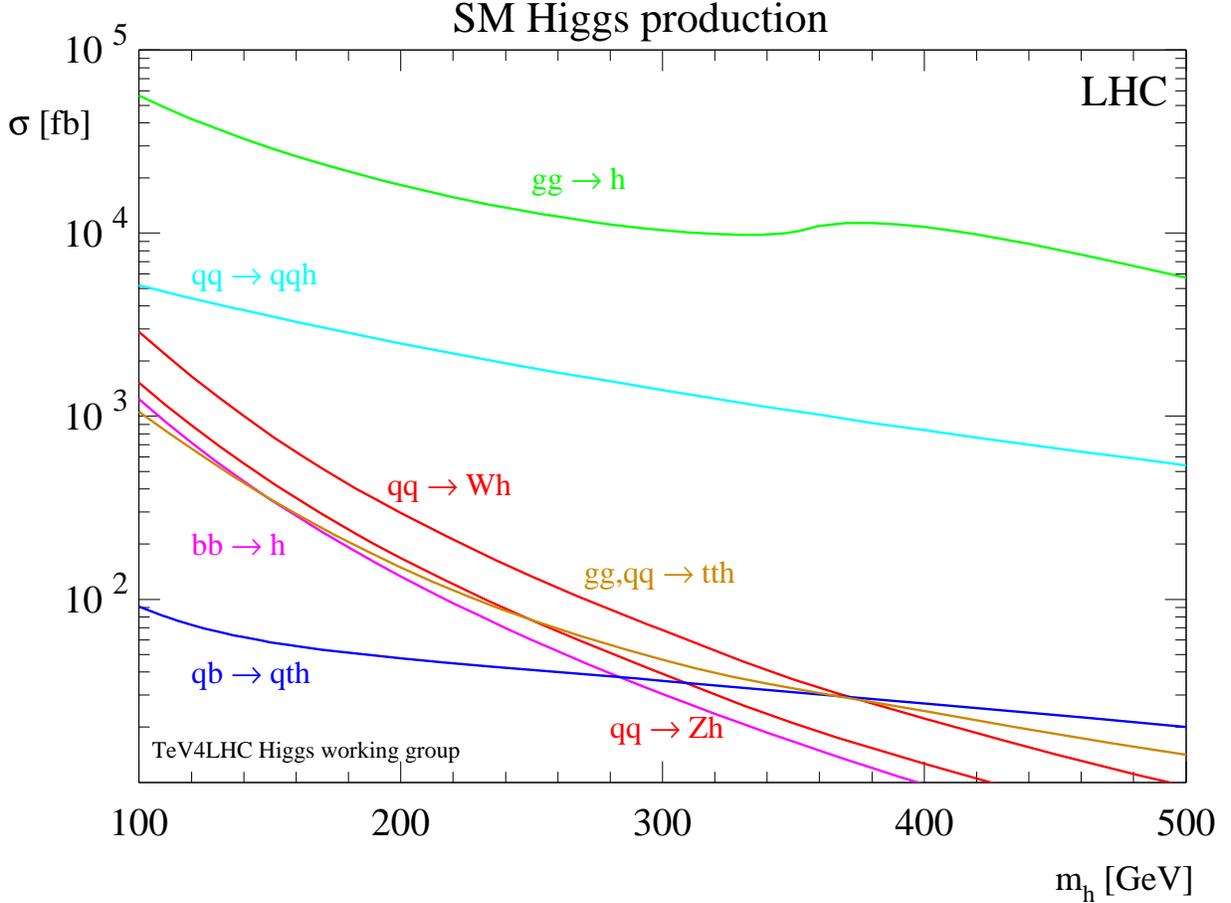}
\end{center}
\caption{Higgs-boson production cross sections (fb) at the LHC
($\sqrt{s}=14 \tev$) for the most relevant production mechanisms as a
function of the Higgs-boson mass. Results for $gg \to h$, 
$q\bar q\to Vh$, $b\bar b \to h$ are at NNLO in the QCD expansion.
Weak boson fusion ($qq \to qq h$) and $t\bar t $ associated
production are at NLO accuracy. Single-top associated production 
($qb \to qth$) is at LO.} 
\label{fig:lhc}
\end{figure}

In this section we collect the predictions for the most important
SM Higgs production processes at the Tevatron and at the LHC. The
relevant cross sections are presented in Figs.~\ref{fig:tev} and 
\ref{fig:lhc} as function of the Higgs mass. The results refer to
fully inclusive cross sections. No acceptance cuts or branching ratios
are applied%
\footnote{
More details and data files can be found at 
{\tt maltoni.web.cern.ch/maltoni/TeV4LHC} .
}%
. We do not consider here diffractive Higgs production,
$pp \to p \oplus H \oplus p$~\cite{diffHSM}. For the discussion of this
channel in the MSSM we refer to \citere{diffHMSSM}. 

We do not aim here at a detailed discussion of the importance of each
signature at the Tevatron or the LHC, but only at providing the most
accurate and up-to-date theoretical predictions. To gauge the progress
made in the last years, it is interesting to compare the accuracy of
the results available in the year 2000, at the time of the Tevatron Higgs
Workshop~\cite{Carena:2000yx}, with those shown here.  All
relevant cross sections are now known at least one order better in the
strong-coupling expansion, and in some cases also electroweak
corrections are available.

\begin{itemize}
\item $gg\to h + X$: gluon fusion \\

This process is known at NNLO in
QCD~\cite{Harlander:2002wh,Anastasiou:2002yz,Ravindran:2003um} (in the 
large top-mass limit) and at NLO in QCD for a quark of an arbitrary
mass circulating in the loop~\cite{Graudenz:1992pv}.
Some N$^3$LO results have recently been obtained in 
\citeres{Moch:2005ky,Laenen:2005uz}.
The NNLO results plotted here are from Ref.~\cite{Catani:2003zt} and
include soft-gluon resummation effects at NNLL. MRST2002 at NNLO has
been used~\cite{Martin:2002aw}, with the renormalization and
factorization scales set 
equal to the Higgs-boson mass. The overall residual theoretical
uncertainty is estimated to be around 10\%. 
The uncertainties due to the large top mass limit approximation 
(beyond Higgs masses of $2 \times m_t$) are difficult to estimate but
expected to be relatively small. Differential results
at NNLO are also available~\cite{Anastasiou:2004xq}.  NLO
(two-loop) EW corrections are known for Higgs masses below
$2m_W$, \cite{Aglietti:2004nj,Degrassi:2004mx}, and range
between 5\% and 8\% of the lowest order term. These EW corrections,
however, are not included in Figs.~\ref{fig:tev}, \ref{fig:lhc}, and
they are also omitted in the MSSM evaluations below.
The same holds for the recent corrections obtained in
\citeres{Moch:2005ky,Laenen:2005uz}.

\item $qq \to qqh + X$: weak boson fusion\\

This process is known at NLO in
QCD~\cite{Han:1992hr,Berger:2004pc,Figy:2003nv}.  Results plotted
here have been obtained with MCFM\cite{Campbell:1999ah}. 
Leading EW corrections are taken into account by using $\al(\MZ)$ as
the (square of the) electromagnetic coupling.
The PDF
used is CTEQ6M~\cite{Pumplin:2002vw} and the renormalization and
factorization scales are 
set equal to the Higgs-boson mass. The theoretical uncertainty is
rather small, less than 10\%.

\item $q \bar q \to Vh + X$: $W,Z$ associated production\\

These processes are known at NNLO in the QCD
expansion~\cite{Brein:2003wg} and at NLO in the electroweak
expansion~\cite{Ciccolini:2003jy}.  The results plotted here have
been obtained by the LH2003 Higgs working group by combining NNLO QCD
and NLO EW corrections~\cite{Assamagan:2004mu}.  The PDF used is
MRST2001 and the renormalization and factorization scales are set
equal to the Higgs-vector-boson invariant mass.  The residual
theoretical uncertainty is rather small, less than 5\%.

\item $b\bar b \to h + X$: bottom fusion\\

This process is known at NNLO in QCD in the five-flavor
scheme~\cite{Harlander:2003ai}.  The cross section in the
four-flavor scheme is known at
NLO~\cite{Dittmaier:2003ej,Dawson:2003kb}.  Results obtained in the
two schemes have been shown to be
consistent~\cite{Assamagan:2004mu,Dawson:2004sh,Dawson:2005vi}. 
The results plotted here are from Ref.~\cite{Harlander:2003ai}.
MRST2002 at NNLO has been used, with the renormalization scale set
equal to $m_h$ and the factorization scale set equal to $m_h/4$.  For
results with one final-state $b$-quark at high-$p_T$ we refer to
\citere{Campbell:2002zm,Dawson:2004sh}.  For results with two
final-state $b$-quarks at high-$p_T$ we refer to
\citere{Dittmaier:2003ej,Dawson:2003kb}.

\item $q\bar q,gg \to t \bar t h + X$: $t \bar t$ associated
       production\\

This process is known at NLO in
QCD~\cite{Beenakker:2001rj,Reina:2001sf,Dawson:2002tg}.  The results
plotted here are from Ref.~\cite{Dawson:2002tg}.  The PDF used is
CTEQ6M and the renormalization and factorization scales are set equal
to $m_t+m_h/2$.

\item $q b \to qth$ : single-top associated production\\

This process is known at LO in QCD~\cite{Ballestrero:1992bk}.  The
results plotted here ($t$-channel production, LHC only) are from
Ref.~\cite{Maltoni:2001hu}. The PDF used is CTEQ5L and the
renormalization and factorization scales are set equal to the
Higgs-boson mass.

\end{itemize}


\section{MSSM Higgs production cross sections}
\label{sec:MSSM}

The MSSM requires two Higgs doublets, resulting in five physical Higgs
boson degrees of freedom.  These are the light and heavy $\cp$-even
Higgs bosons, $h$ and $H$, the $\cp$-odd Higgs boson, $A$, and the
charged Higgs boson, $H^\pm$.  The Higgs sector of the MSSM can be
specified at lowest order in terms of $\MZ$, $\MA$, and $\tb \equiv
v_2/v_1$, the ratio of the two Higgs vacuum expectation values.  The
masses of the $\cp$-even neutral Higgs bosons and the charged Higgs
boson can be calculated, including higher-order corrections, in terms
of the other MSSM parameters.

After the termination of LEP in the year 2000 (the final LEP
results can be found in \citeres{LEPHiggsSM,LEPHiggsMSSM}), the Higgs
boson search has shifted to the Tevatron and will later be continued
at the LHC.  For these analyses and investigations a precise
prediction of the Higgs boson masses, branching ratios and production
cross sections in the various channels is necessary.

Due to the large number of free parameters, a complete scan of the
MSSM parameter space is too involved. Therefore the search results at
LEP~\cite{LEPHiggsMSSM} and the
Tevatron~\cite{D0bounds,CDFbounds,Tevcharged}, as well as studies for the
LHC~\cite{schumi} have been performed in several benchmark
scenarios~\cite{benchmark,benchmark2,benchmark3}.

The code {\tt FeynHiggs}~\cite{feynhiggs,mhiggsletter,mhiggslong,mhiggsAEC}
provides a precise calculation of the Higgs boson mass spectrum, couplings and
the decay widths%
\footnote{
The code can be obtained from {\tt www.feynhiggs.de} .
}%
. This has now been supplemented by the evaluation of
all relevant neutral Higgs boson production cross sections at the
Tevatron and the LHC (and the corresponding three SM cross sections
for both colliders with $M_H^{\rm SM} = \mh, m_H, m_A$). 
They are calculated by using the effective
coupling approach, rescaling the SM result%
\footnote{
The inclusion of the charged Higgs production cross sections is
planned for the near future.
}%
.

In this section we will briefly describe the benchmark scenarios with
their respective features. The effective coupling approach, used to
obtain the production cross sections within {\tt FeynHiggs}, is
discussed. Results for the neutral Higgs production cross sections at
the Tevatron and the LHC are presented within the benchmark scenarios
for two values of $\tb$, $\tb = 5, 40$.

\subsection{The benchmark scenarios}
\label{ssec:bench}

We start by recalling the four benchmark scenarios~\cite{benchmark2}
suitable for the MSSM Higgs boson search at hadron colliders%
\footnote{
In the course of this workshop they have been refined to cover wider
parts of the MSSM parameter space relevant especially for heavy MSSM
Higgs boson production~\cite{benchmark3}.
}%
. 
In these scenarios the values of the parameters of the $\Stop$~and
$\Sbot$~sector as well as the gaugino masses are fixed, while
$\tb$ and $\MA$ are the parameters that are varied. Here we fix $\tb$
to a low and a high value, $\tb = 5, 40$, but vary $\MA$. This also
yields a variation of $\Mh$ and $\MH$.

In order to fix our notations, we list the conventions for the inputs
from the scalar top and scalar bottom sector of the MSSM:
the mass matrices in the basis of the current eigenstates $\StopL, \StopR$ and
$\SbotL, \SbotR$ are given by
\BEA
\label{stopmassmatrix}
{\cal M}^2_{\Stop} &=&
  \ML \MstL^2 + \mt^2 + \CZb (\edz - \frac{2}{3} \sw^2) \MZ^2 &
      \mt \Xt \\
      \mt \Xt &
      \MstR^2 + \mt^2 + \frac{2}{3} \CZb \sw^2 \MZ^2 
  \MR, \\
&& \non \\
\label{sbotmassmatrix}
{\cal M}^2_{\Sbot} &=&
  \ML \MsbL^2 + \mb^2 + \CZb (-\edz + \frac{1}{3} \sw^2) \MZ^2 &
      \mb \Xb \\
      \mb \Xb &
      \MsbR^2 + \mb^2 - \frac{1}{3} \CZb \sw^2 \MZ^2 
  \MR,
\EEA
where 
\BE
\mt \Xt = \mt (\At - \mu \CTb) , \quad
\mb\, \Xb = \mb\, (\Ab - \mu \tb) .
\label{eq:mtlr}
\EE
Here $\At$ denotes the trilinear Higgs--stop coupling, $\Ab$ denotes the
Higgs--sbottom coupling, and $\mu$ is the higgsino mass parameter.
SU(2) gauge invariance leads to the relation
\BE
\MstL = \MsbL .
\EE
For the numerical evaluation, a convenient choice is
\BE
\MstL = \MsbL = \MstR = \MsbR =: \msusy .
\label{eq:msusy}
\EE
The parameters in the $\Stop/\Sbot$ sector are defined here as
on-shell parameters, see \citere{bse} for a discussion and a
translation to \drbar\ parameters.
The top-quark mass is taken to be  
$\mt = \mt^{\rm exp} = 172.7 \gev$~\cite{mt1727}.

\begin{itemize}
\item{The $\mhmax$ scenario:}

This scenario had been designed to obtain conservative $\tb$ exclusion
bounds~\cite{tbexcl}.
The parameters are chosen such that the maximum possible 
Higgs-boson mass as a function of $\tb$ is obtained
(for fixed $\msusy$ and $\mt$, 
and $\MA$ set to its maximal value, $\MA = 1 \tev$).
The parameters are%
\footnote{
As mentioned above, no external constraints are taken into account.
In the minimal flavor violation scenario,  better agreement with
${\rm BR}(b \to s \gamma)$ constraints would be 
obtained for the other sign of $\Xt$ (called the
``constrained $\mhmax$'' scenario~\cite{benchmark2}). 
}%
:
\BEA
&& \msusy = 1 \tev, \; 
\mu = 200 \gev, \;
M_2 = 200 \gev, \non \\
&& \Xt = 2\, \msusy  \; 
\Ab = \At, 
\mgl = 0.8\,\msusy~.
\label{mhmax}
\EEA

\item{The no-mixing scenario:}

This benchmark scenario is associated with
vanishing mixing in the $\Stop$~sector and with a higher SUSY mass
scale as compared to the $\mhmax$~scenario to increase the parameter
space that avoids the LEP Higgs bounds:
\BEA
&& \msusy = 2 \tev, \; 
\mu = 200 \gev, \;
M_2 = 200 \gev, \non \\
&& \Xt = 2\, \msusy  \; 
\Ab = \At, 
\mgl = 0.8\,\msusy~.
\label{nomix}
\EEA

\item{The gluophobic Higgs scenario:}

In this scenario the main production cross section for the light Higgs
boson at the LHC, $gg \to h$, can be strongly suppressed for a wide range
of the $\MA-\tb$-plane. This happens due to a
cancellation between the top quark and the stop quark loops in the
production vertex (see \citere{ggsuppr}). This cancellation is more
effective for small $\Stop$~masses and for relatively large
values of the $\Stop$~mixing parameter, $\Xt$. The partial width of
the most relevant decay mode, $\Gamma(h \to \gamma\gamma)$, is affected much
less, since it is dominated by the $W$~boson loop.
The parameters are:
\BEA
&& \msusy = 350 \gev, \; 
\mu = 300 \gev, \;
M_2 = 300 \gev, \non \\
&& \Xt = -750 \gev \; 
\Ab = \At, 
\mgl = 500 \gev~.
\label{ggsup}
\EEA

\item{The small $\aeff$ scenario:}

Besides the channel $gg \to h \to \ga\ga$ at the LHC, the other
channels for light Higgs 
searches at the Tevatron and at the LHC mostly rely on the decays 
$h \to b \bar b$ and $h \to \tau^+\tau^-$. Including Higgs-propagator
corrections the couplings 
of the lightest Higgs boson to down-type fermions are $\sim
\sin\aeff$, where $\aeff$ is the loop corrected mixing angle in the
neutral $\cp$-even Higgs sector.
Thus, if $\aeff$ is small, the two main decay channels can be heavily
suppressed in the MSSM compared to the SM case. 
Such a suppression occurs for large $\tb$ and not too large $\MA$.
The parameters of this scenario are:
\BEA
&& \msusy = 800 \gev, \;
\mu = 2.5 \, \msusy, \;
M_2 = 500 \gev, \non \\
&& \Xt = -1100 \gev, \; 
\Ab = \At, 
\mgl = 500 \gev~.
\label{smallaeff}
\EEA

\end{itemize}


\subsection{The effective coupling approximation}
\label{ssec:eff}

We consider the following neutral Higgs production cross sections at
the Tevatron and the LHC ($\phi$ denotes all neutral MSSM Higgs
bosons, $\phi = h, H, A$):
\BEA
gg &\to& \phi + X~, \\
qq &\to& qq\phi + X~, \\
q \bar q &\to& W/Z\phi + X~, \\
b \bar b &\to& \phi + X~, \\
gg, qq &\to& t \bar t \phi~.
\EEA
The MSSM cross sections have been obtained by rescaling the corresponding 
SM cross sections of Section~\ref{sec:SM} either with the ratio of the
corresponding MSSM decay with 
(of the inverse process) over the SM decay width, or with the square
of the ratio of the corresponding couplings.
More precisely, we apply the following factors:
\begin{itemize}

\item{$gg \to \phi + X$:}
\BE
\frac{\Ga(\phi \to gg)_{\MSSM}}{\Ga(\phi \to gg)_{\SM}}
\label{eq:phigg}
\EE
We include the full one-loop result with SM QCD corrections. MSSM
two-loop corrections~\cite{gghMSSM2L} have been neglected.

\item{$qq \to qq\phi + X$:}
\BE
\frac{|g_{\phi VV, \MSSM}|^2}{|g_{\phi VV, \SM}|^2} , \; V = W, Z~.
\EE
We include the full set of Higgs propagator corrections in the
effective couplings.

\item{$qq \to W/Z\phi + X$:}
\BE
\frac{|g_{ \phi VV, \MSSM}|^2}{|g_{\phi VV, \SM}|^2} , \; V = W, Z~.
\EE
We include the full set of Higgs propagator corrections in the
effective couplings.

\item{$b \bar b \to \phi + X$:}
\BE
\frac{\Ga(\phi \to b \bar b)_{\MSSM}}{\Ga(\phi \to b \bar b)_{\SM}}~.
\EE
We include here one-loop SM QCD and SUSY QCD corrections, as well as
the resummation of all terms of \order{(\al_s \tb)^n}.

\item{$gg, qq \to t \bar t \phi$:}
\BE
\frac{|g_{\phi t \bar t, \MSSM}|^2}{|g_{\phi t \bar t, \SM}|^2}~,
\label{eq:phitt}
\EE
where $g_{\phi t \bar t, \MSSM}$ and $g_{\phi t \bar t, \SM}$ are
composed of a left- and a right-handed part.
We include the full set of Higgs propagator corrections in the
effective couplings.

\end{itemize}

In the effective couplings introduced in
eqs.~(\ref{eq:phigg})--(\ref{eq:phitt}) we have used the proper
normalization of the external (on-shell) Higgs bosons as discussed in
\citere{Hahn:2002gm}.

It should be noted that the effective coupling approximation as
described above does not take into account the MSSM-specific
dynamics of the production processes. The theoretical uncertainty in
the predictions for the cross sections will therefore in general be
somewhat larger than for the decay widths.


\subsection{Results}

Results for the neutral Higgs production cross sections at
the Tevatron and the LHC are presented within the four benchmark scenarios
for two values of $\tb$, $\tb = 5, 40$, giving a total of eight plots
for each collider. 
 
\begin{figure}[p]
\begin{center}
\includegraphics[width=8cm]{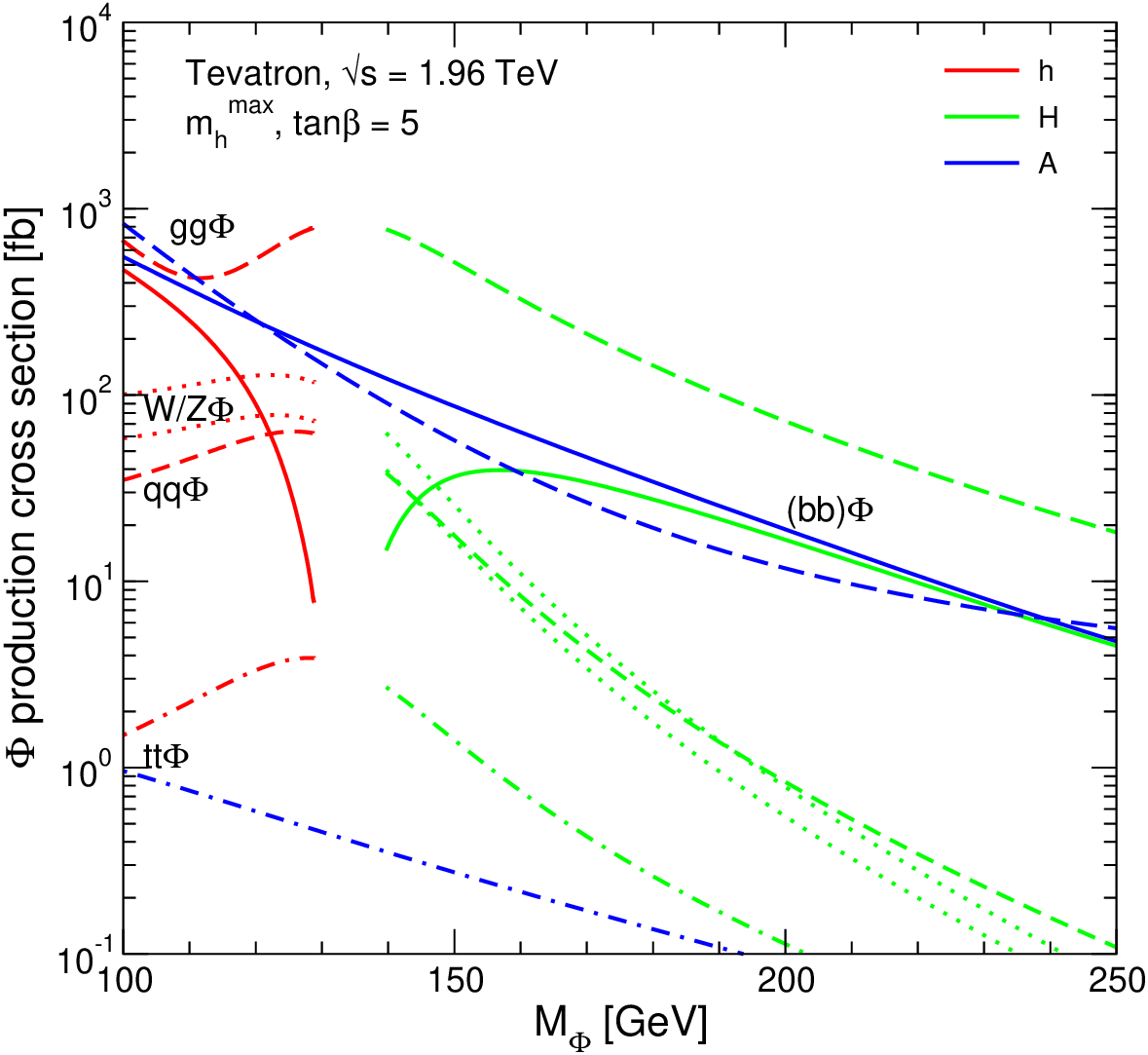}
\includegraphics[width=8cm]{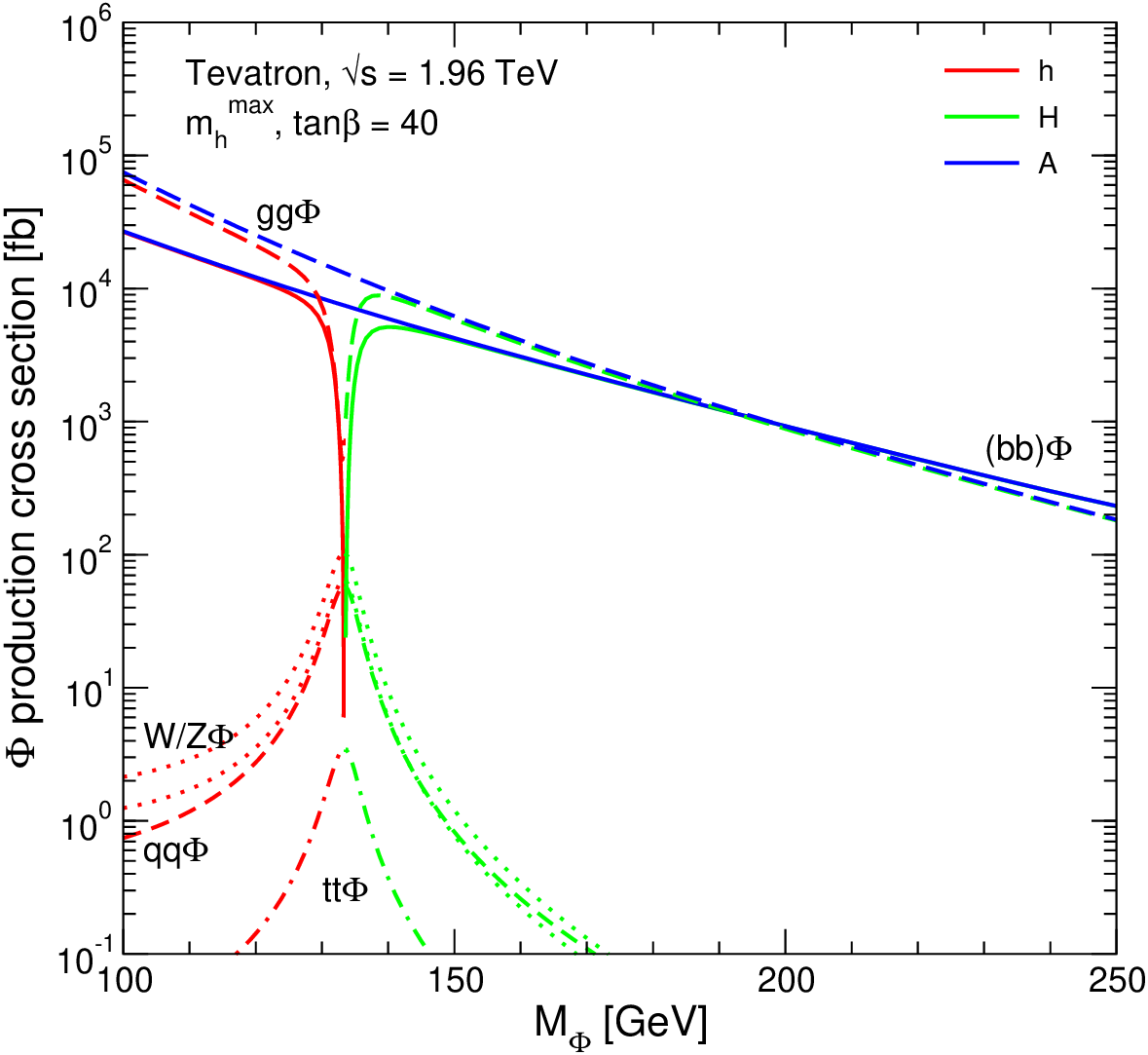}\\
\includegraphics[width=8cm]{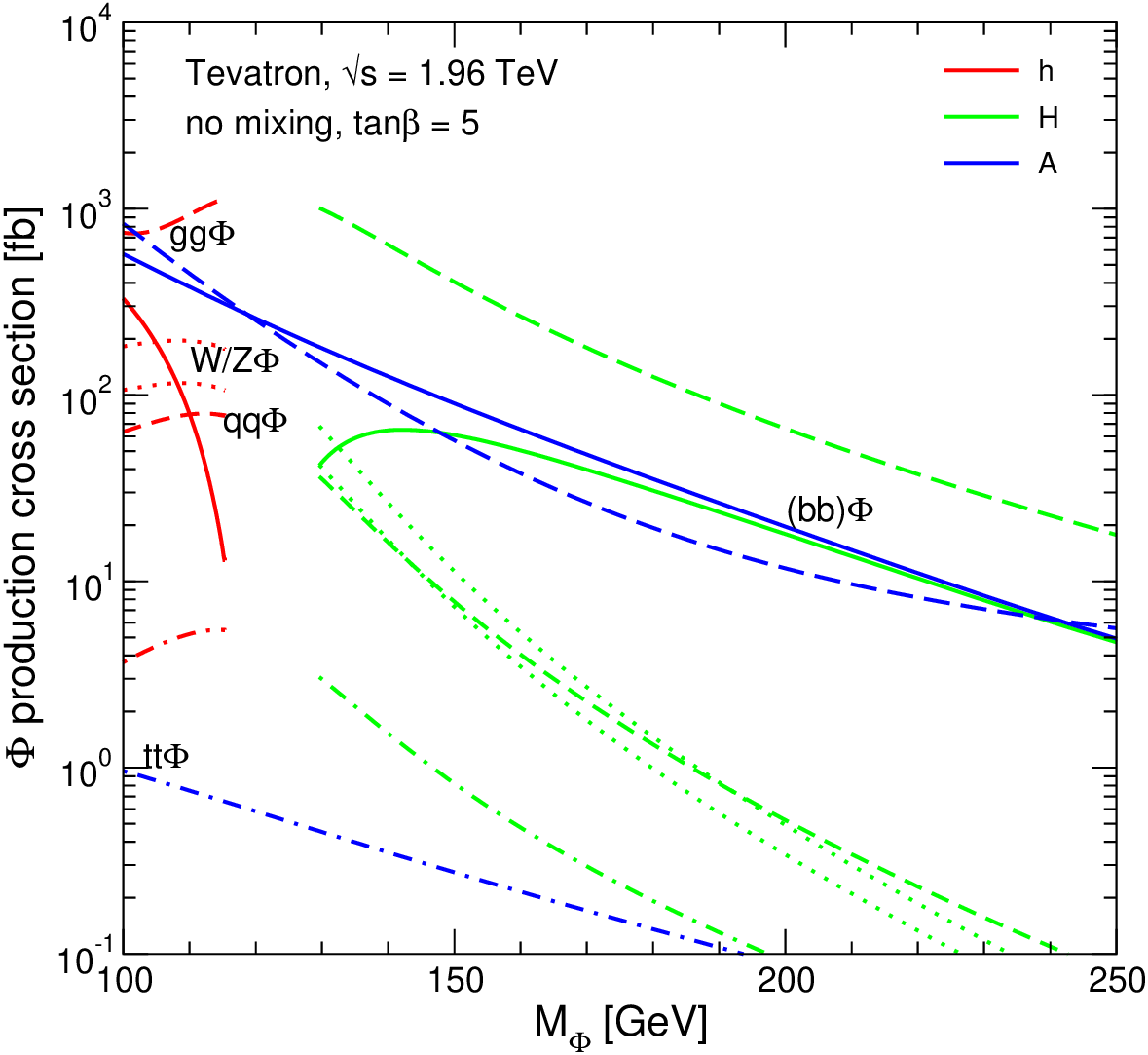}
\includegraphics[width=8cm]{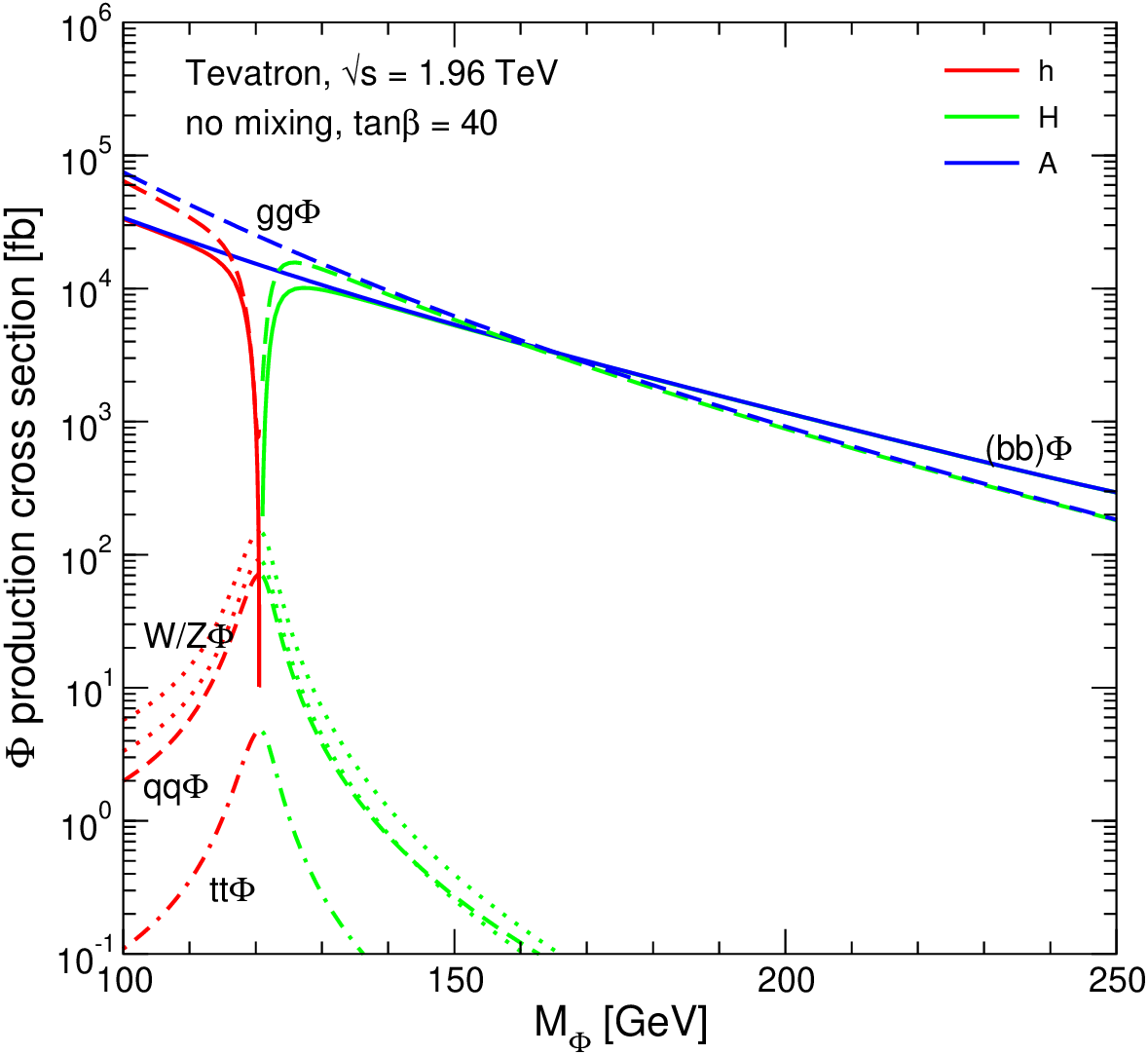}\\
\end{center}
\caption{Neutral Higgses production cross sections (fb) at the
Tevatron ($\sqrt{s}=1.96 \tev$) for the most relevant production
mechanisms as a function of the Higgs-boson mass. Results 
are based on the SM cross sections and evaluated through an effective
coupling approximation in the $\mhmax$ and no-mixing scenarios, for
$\tb=5,40$ and $\Phi = h, H, A$.} 
\label{fig:mssm_tev_1}
\end{figure}

\begin{figure}[p]
\begin{center}
\includegraphics[width=8cm]{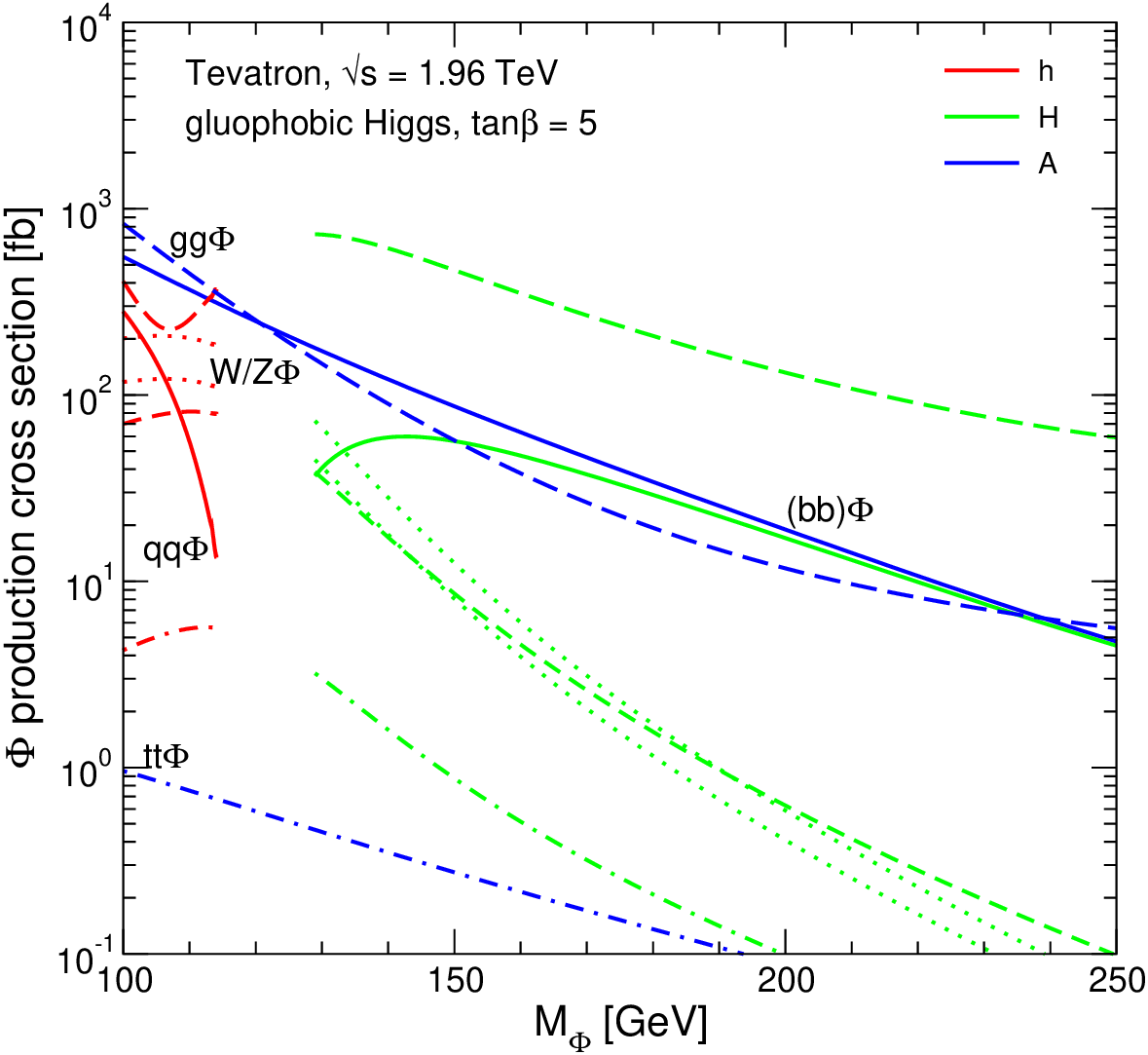}
\includegraphics[width=8cm]{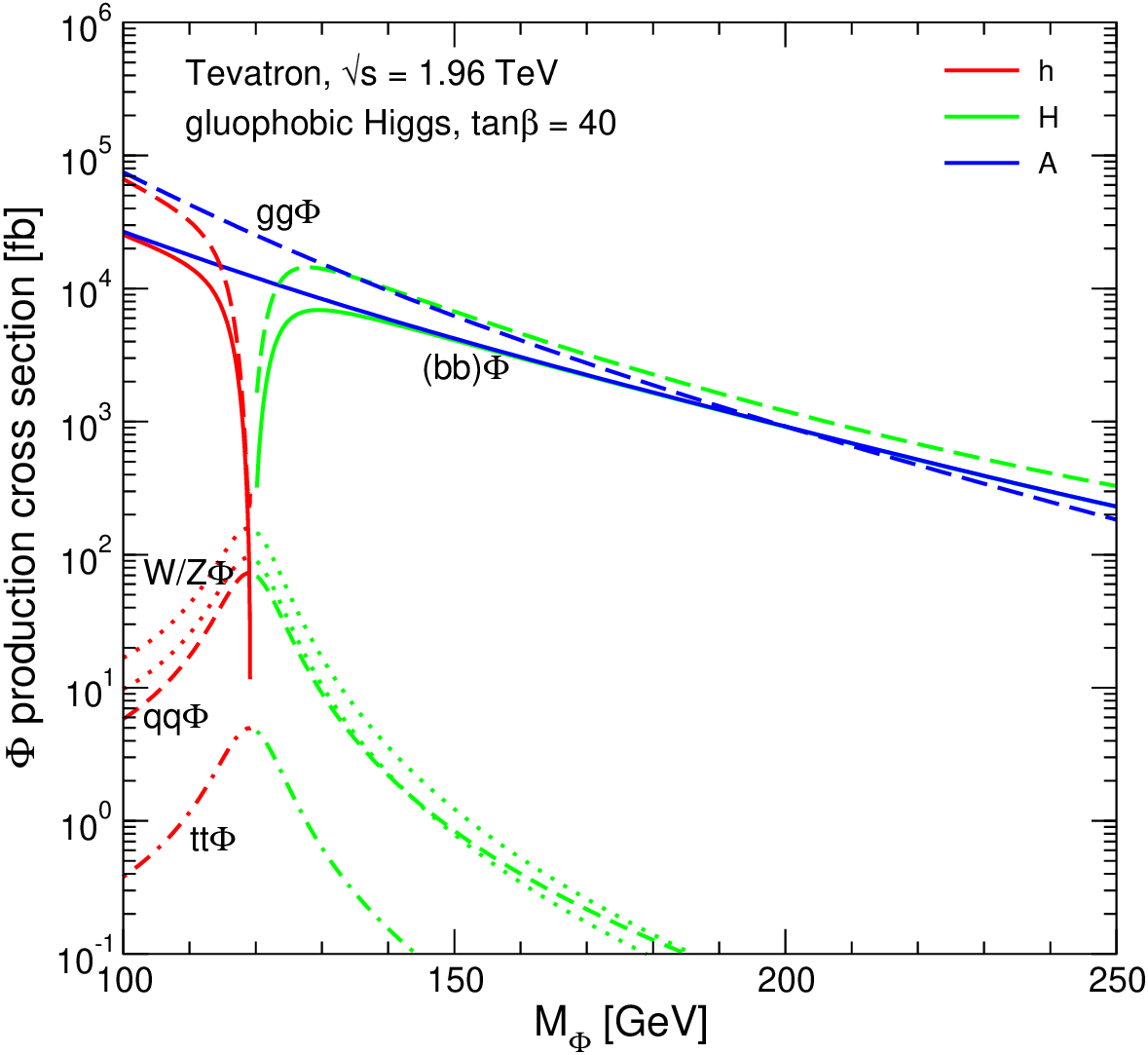}
\includegraphics[width=8cm]{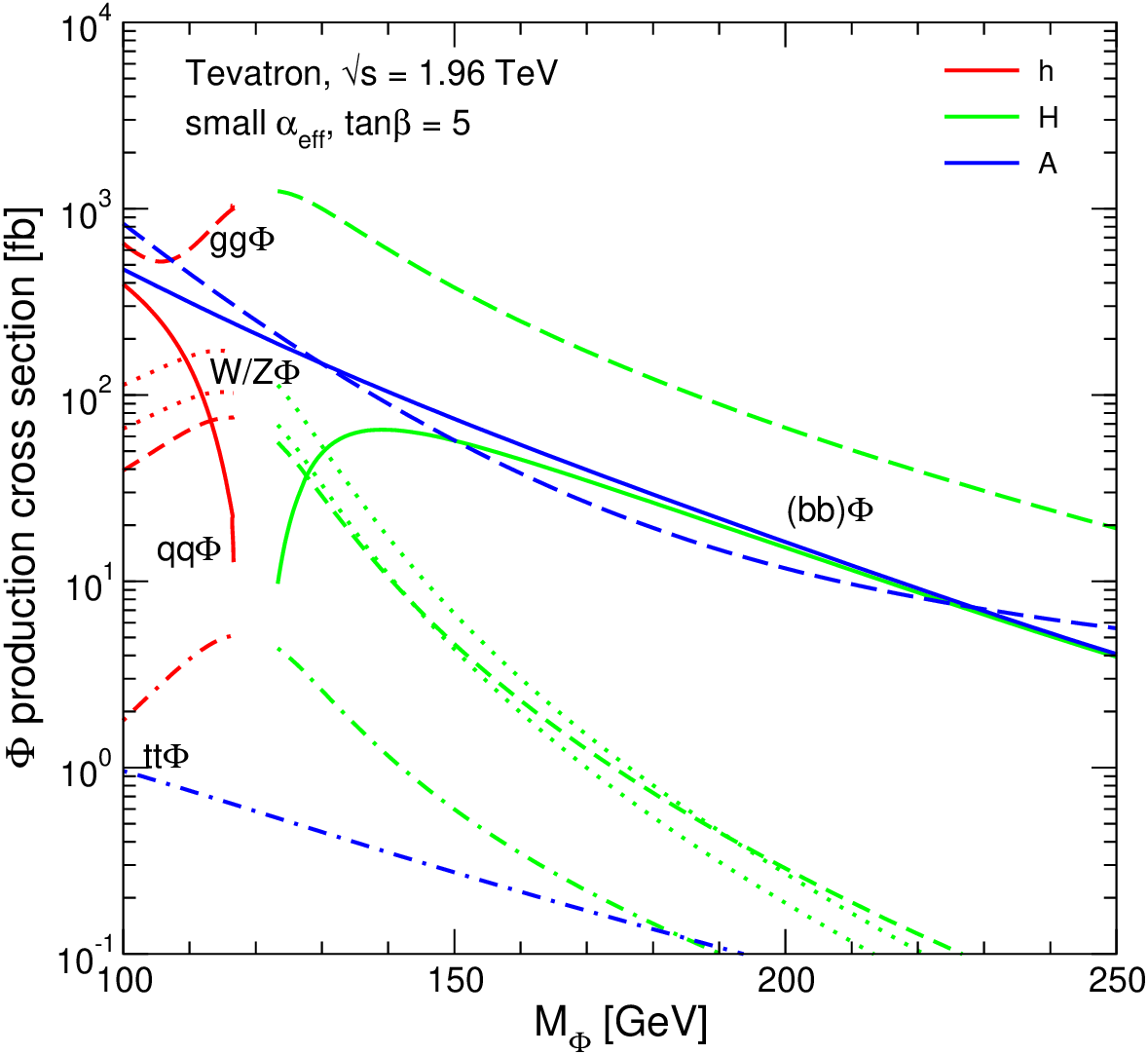}
\includegraphics[width=8cm]{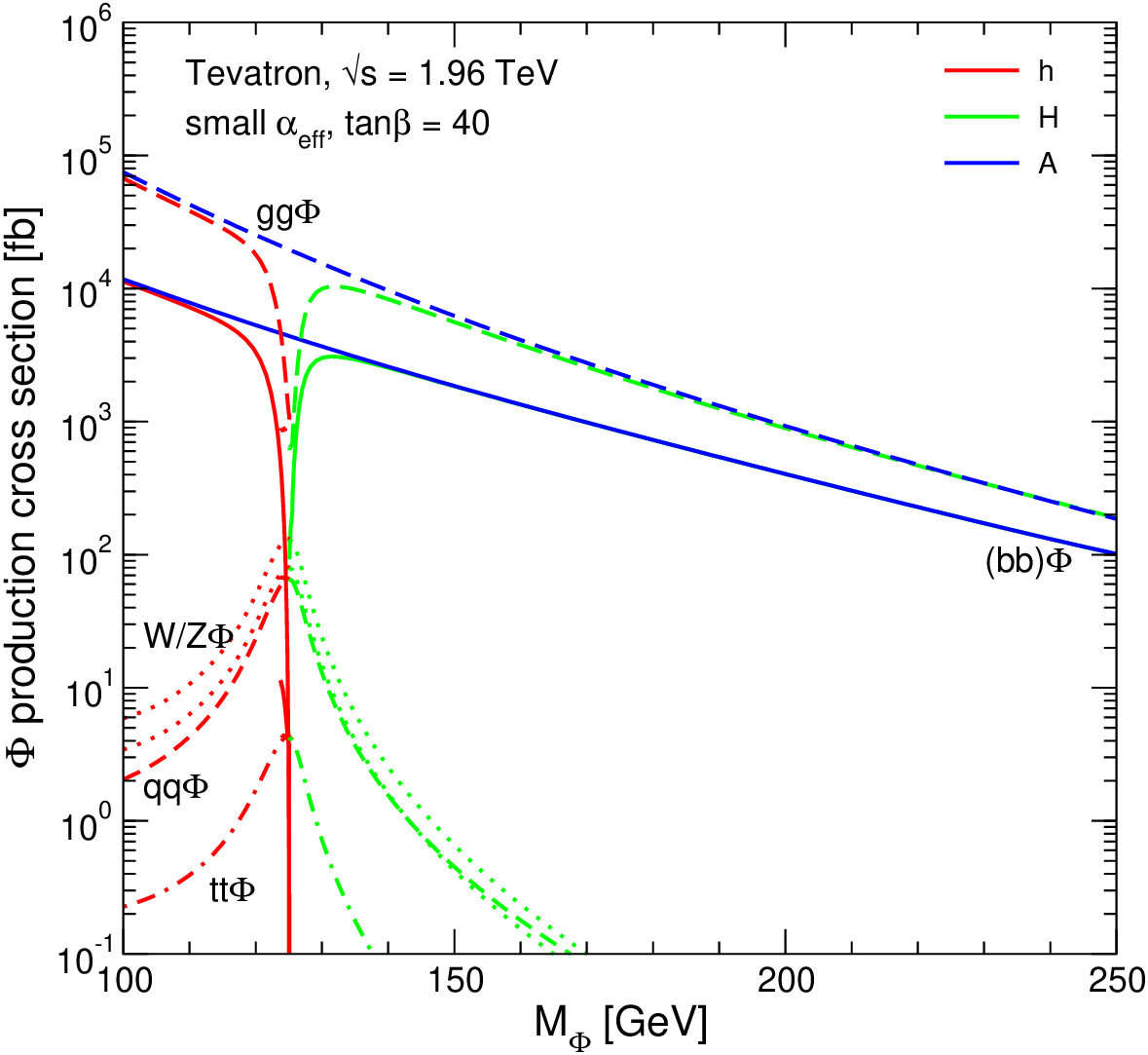}\\
\end{center}
\caption{Same as Fig.~\ref{fig:mssm_tev_1}, for the gluophobic Higgs
and small $\aeff$ scenarios.} 
\label{fig:mssm_tev_2}
\end{figure}

\begin{figure}[p]
\begin{center}
\includegraphics[width=8cm]{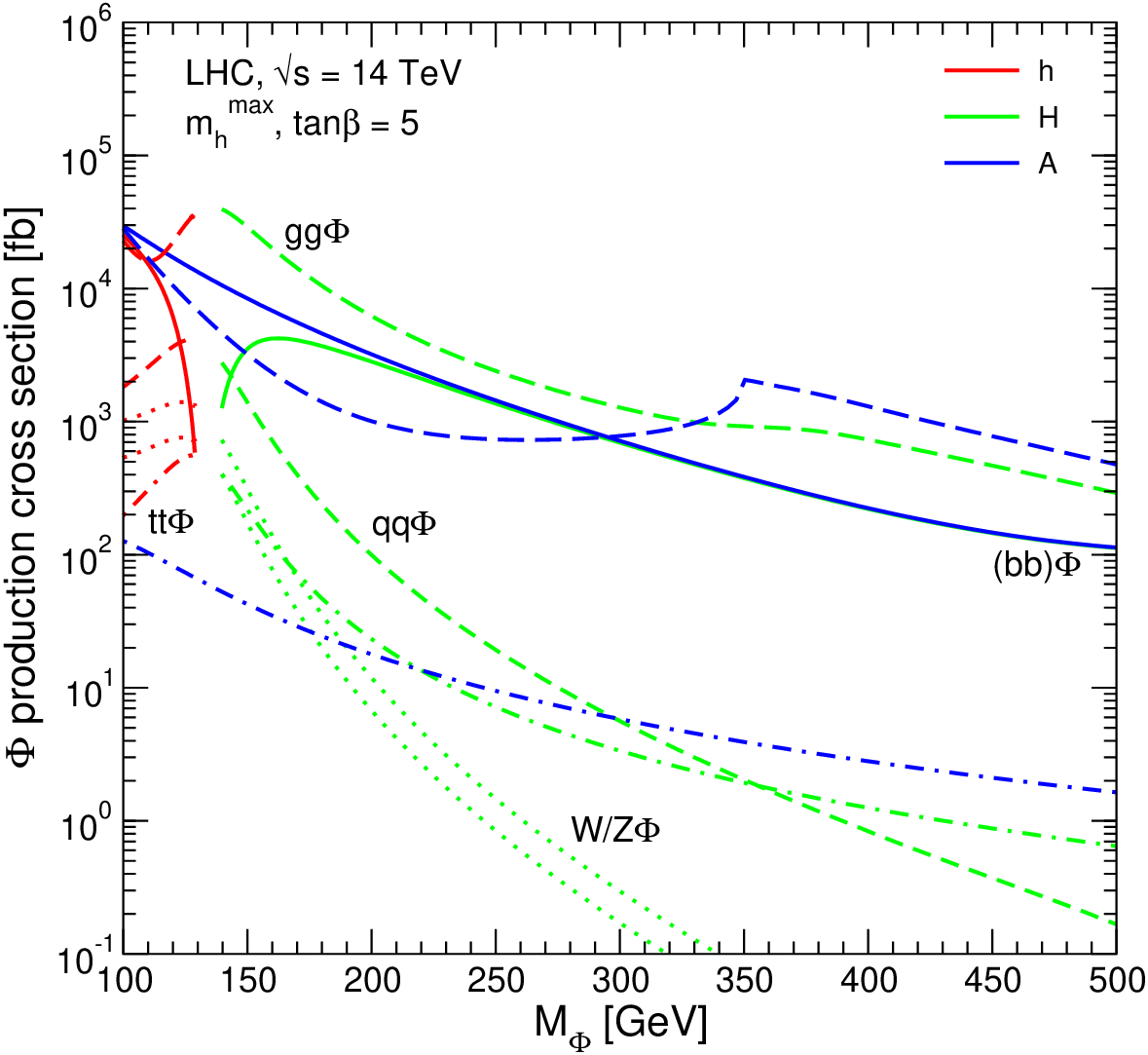}
\includegraphics[width=8cm]{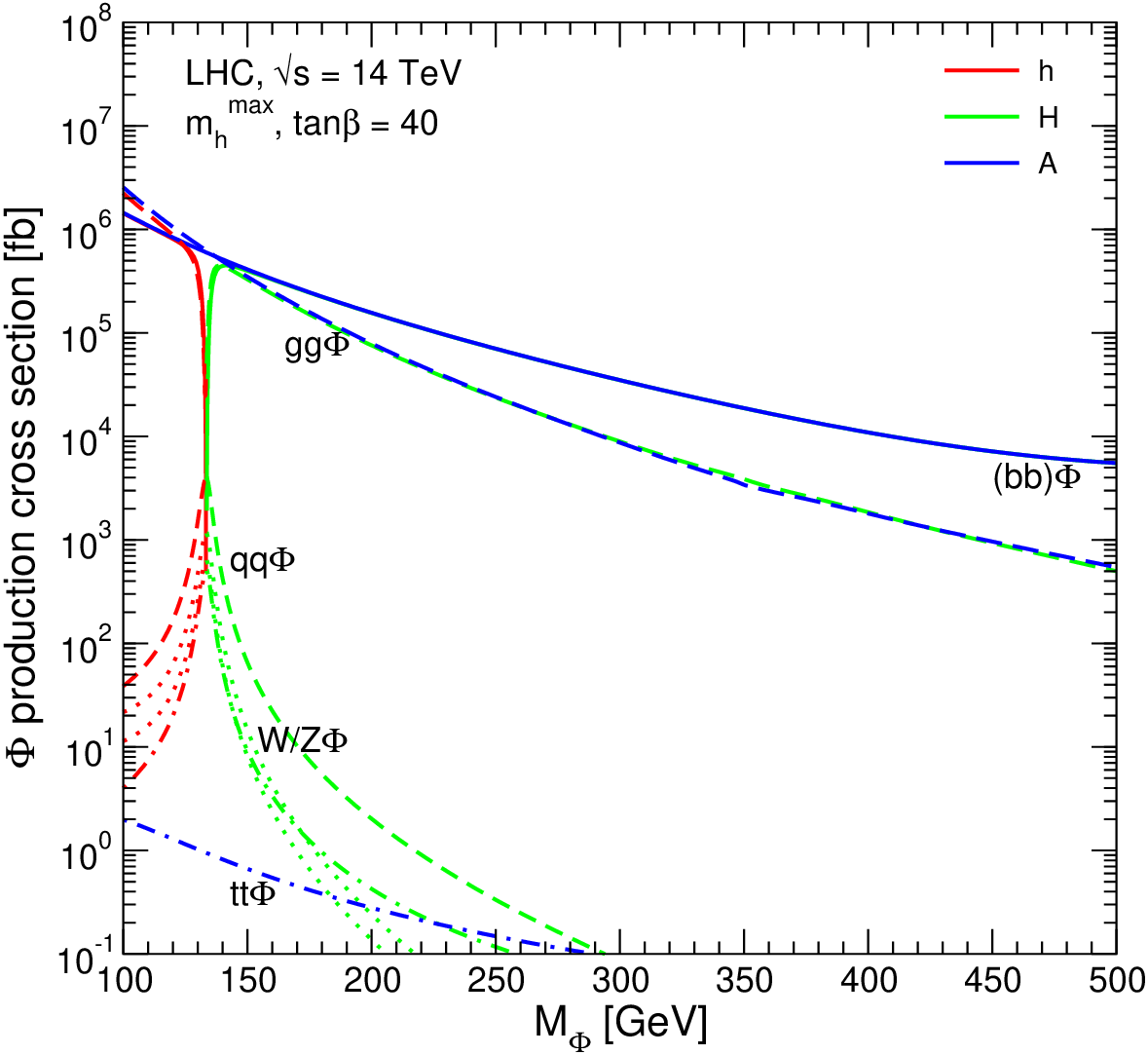}\\
\includegraphics[width=8cm]{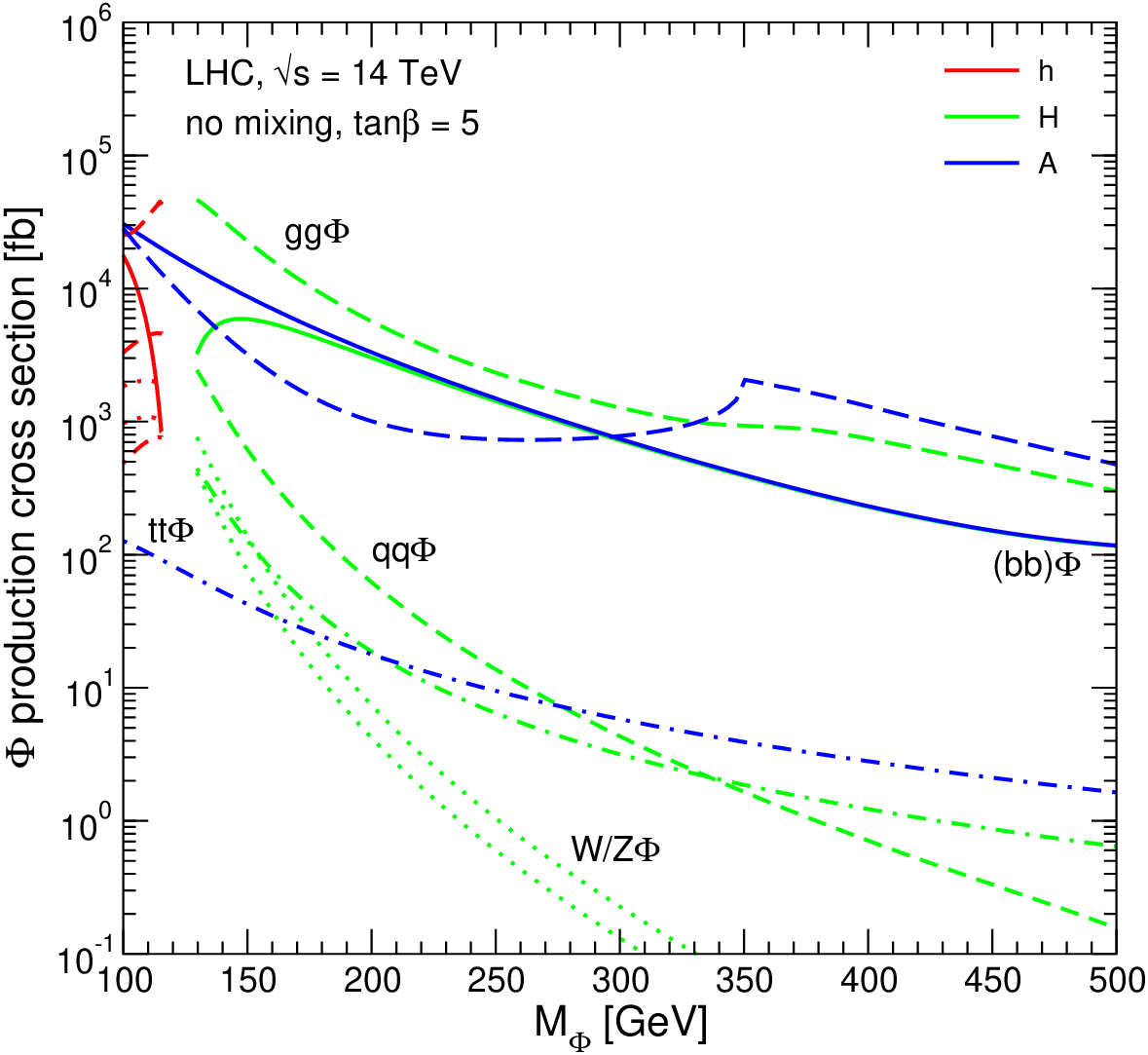}
\includegraphics[width=8cm]{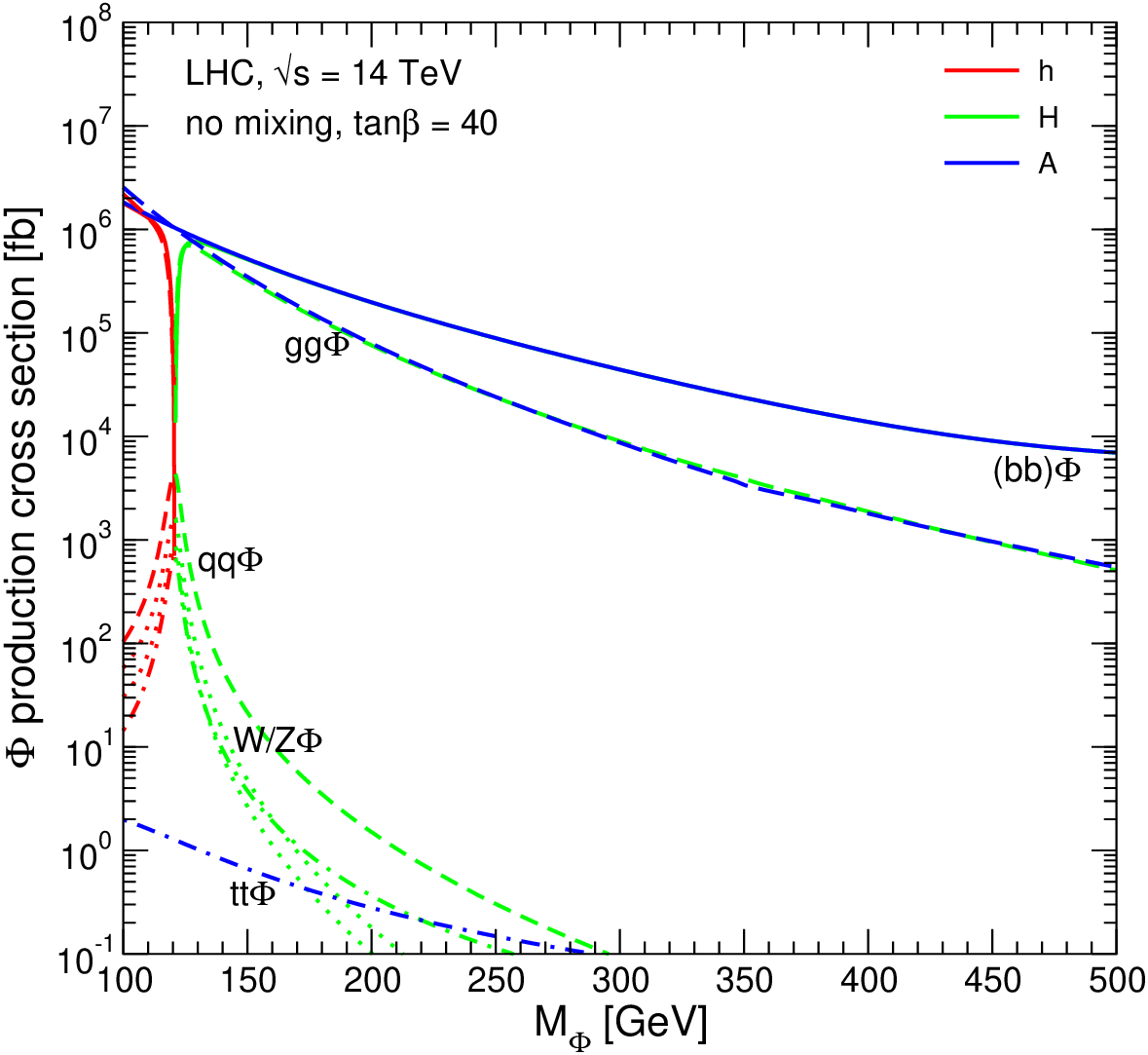}\\
\end{center}
\caption{Neutral Higgses production cross sections (fb) at the LHC
($\sqrt{s}=14 \tev$)
for the most relevant production mechanisms as a function of the
Higgs-boson mass. Results 
are based on the SM cross sections and evaluated through an effective
coupling approximation 
in the $\mhmax$ and no-mixing scenarios, for $\tb=5,40$ and $\Phi = h, H, A$.}
\label{fig:mssm_lhc_1}
\end{figure}

\begin{figure}[p]
\begin{center}
\includegraphics[width=8cm]{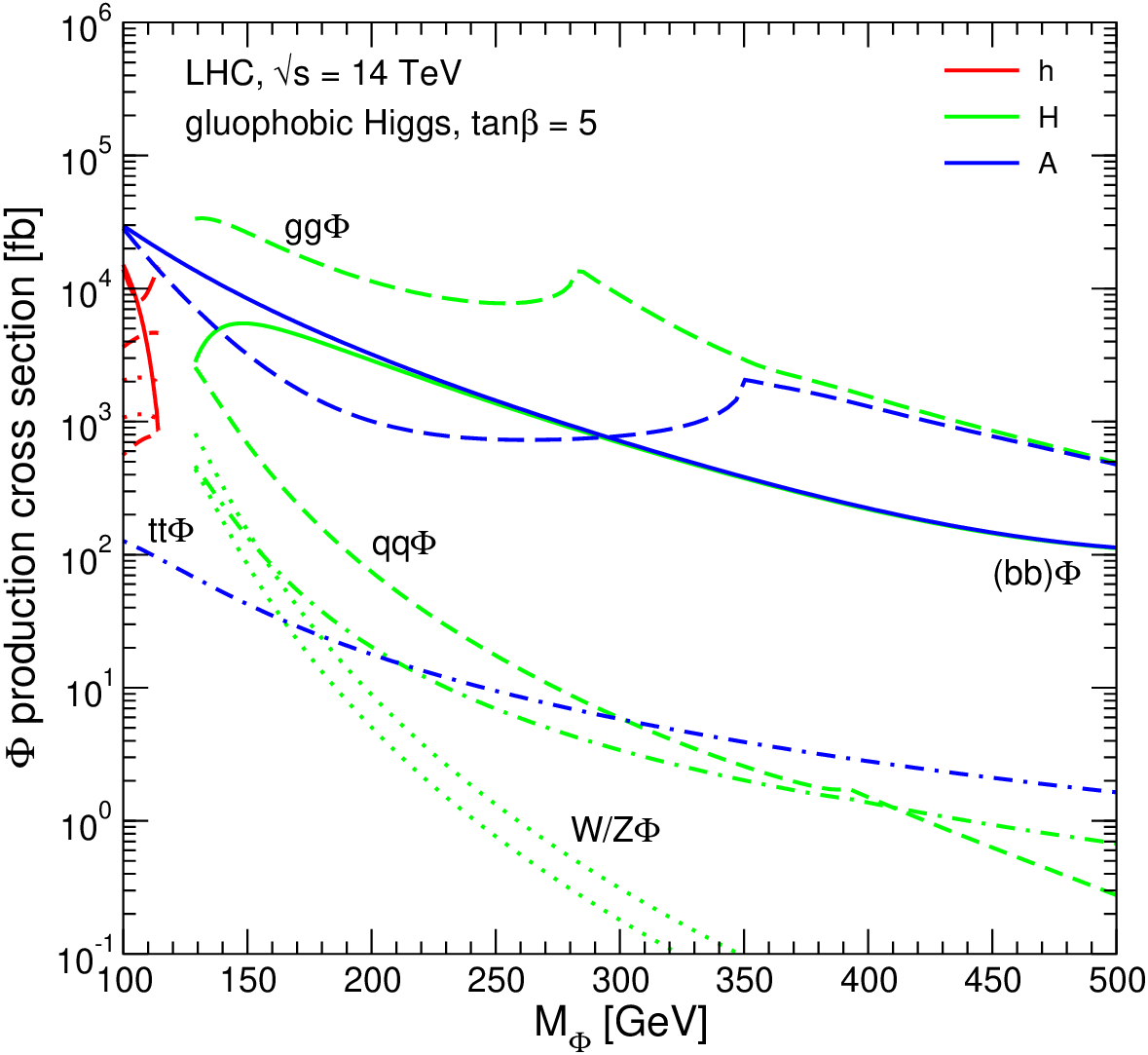}
\includegraphics[width=8cm]{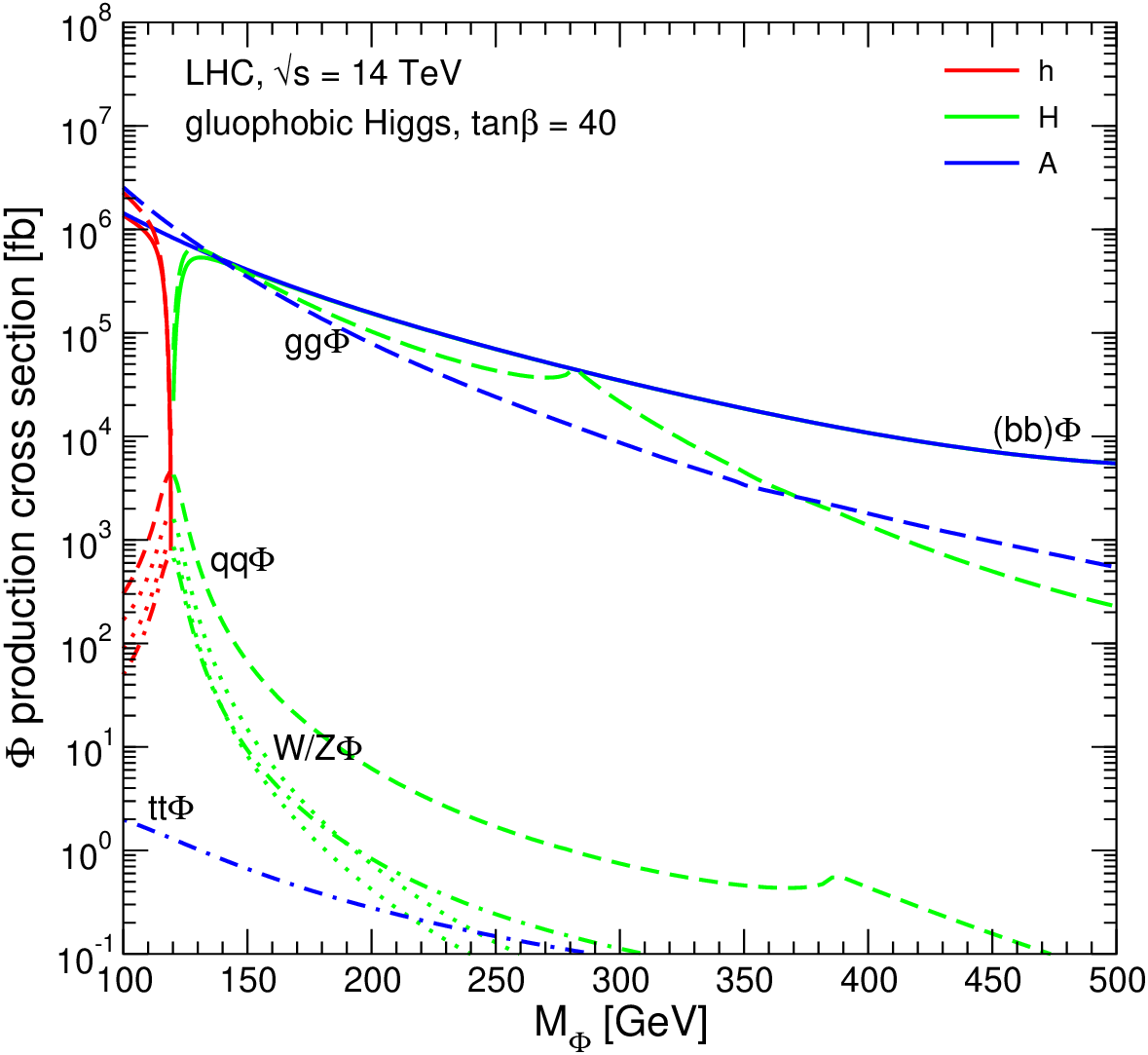}
\includegraphics[width=8cm]{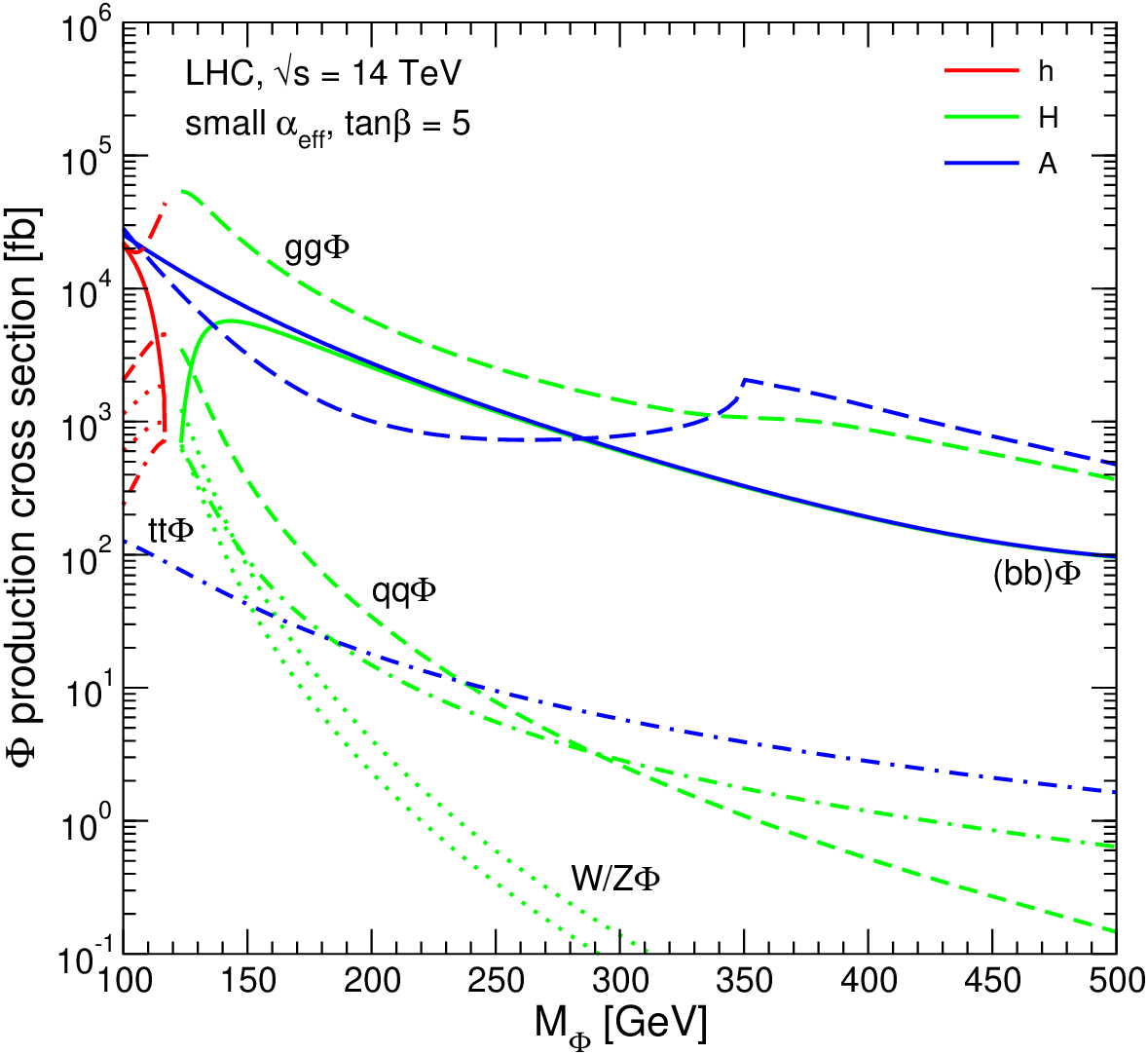}
\includegraphics[width=8cm]{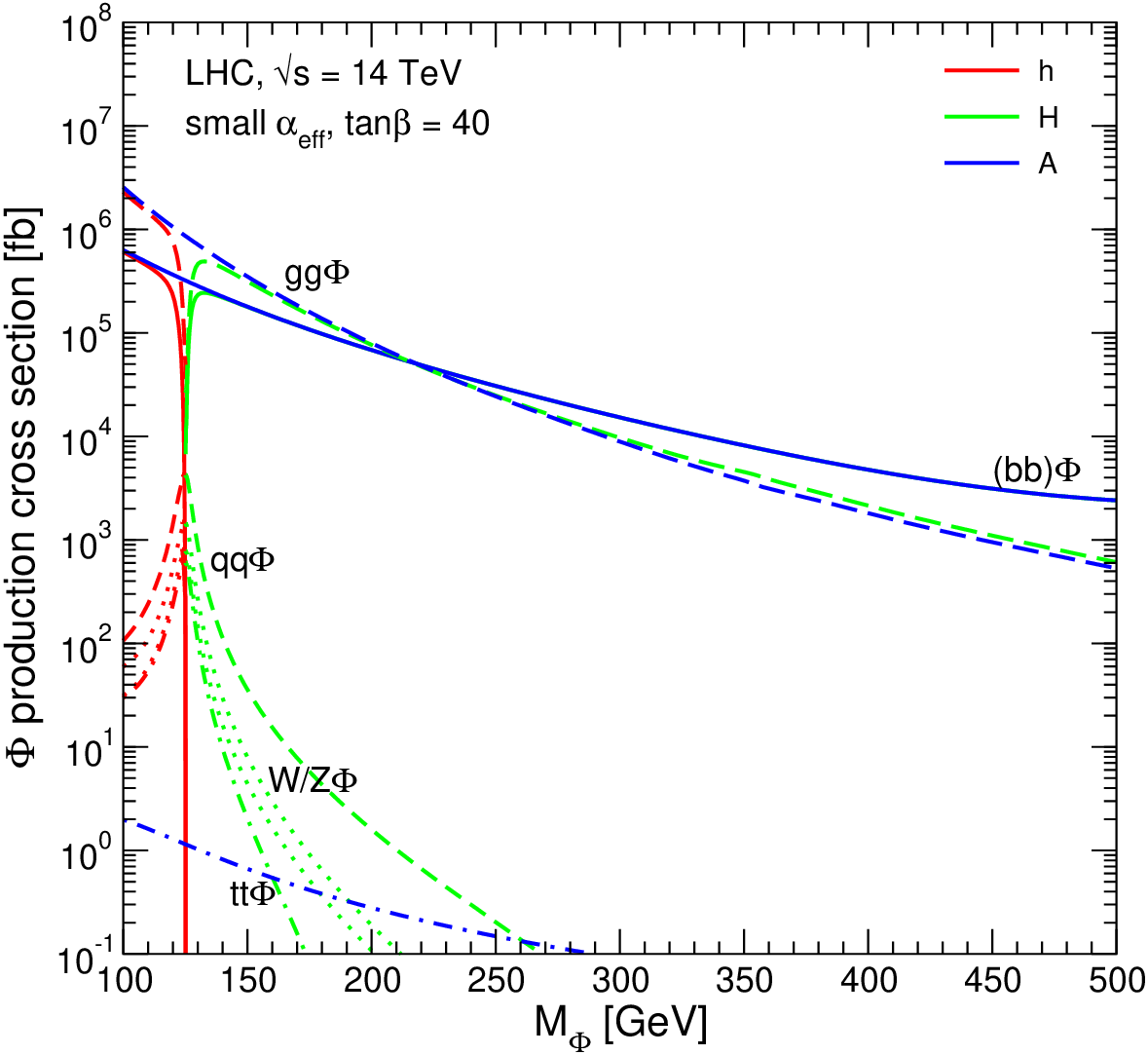}\\
\end{center}
\caption{Same as Fig.~\ref{fig:mssm_lhc_1}, for the gluophobic Higgs
and small $\aeff$ scenarios.} 
\label{fig:mssm_lhc_2}
\end{figure}

Figs.~\ref{fig:mssm_tev_1} and~\ref{fig:mssm_tev_2}  
show the results for the Tevatron, while Figs.~\ref{fig:mssm_lhc_1} 
and~\ref{fig:mssm_lhc_2} show
the LHC results. In Fig.~\ref{fig:mssm_tev_1} (\ref{fig:mssm_lhc_1}) 
the Higgs production cross sections for the
neutral MSSM Higgs bosons at the Tevatron (LHC) in the $m_h^{\rm max}$
scenario (upper row) and the no-mixing scenario (lower row) can be found.
Fig.~\ref{fig:mssm_tev_2} (\ref{fig:mssm_lhc_2}) 
depicts the same for the gluophobic Higgs scenario (upper row)
and the small $\alpha_{\rm eff}$ scenario (lower row). 

For low $M_A$ values the production cross section of the $h$ and the $A$ 
are similar, while for large $M_A$ the cross sections of $H$ and $A$ 
are very close. This effect is even more pronounced for large $\tan\beta$.

The results presented in this paper have been obtained for the MSSM
with real parameters, i.e.\ the $\cp$-conserving case. They can can
easily be extended via the effective coupling approximation to the
case of non-vanishing complex phases (as implemented in {\tt FeynHiggs}).

\section*{Acknowledgements}
We are thankful to John Campbell, Mariano Ciccolini, Massimiliano Grazzini, 
Robert Harlander and Michael Kr\"amer for making some SM predictions
available to us.  
F.M.\ thanks Alessandro Vicini for useful discussions. 
S.H., F.M.\ and G.W.\ thank Michael Spira for lively
discussions.
S.H.\ is partially
supported by CICYT (grant FPA2004-02948) and DGIID-DGA
(grant 2005-E24/2).

\newpage
\bibliographystyle{plain}

\begin{thebibliography}{99}


\bibitem{LEPEWWG} [The ALEPH, DELPHI, L3, OPAL, SLD Collaborations,
                  the LEP Electroweak Working Group,
                  the SLD Electroweak and Heavy Flavour Groups],
                  hep-ex/0509008;\\ \mbox{}
                  [The ALEPH, DELPHI, L3 and OPAL Collaborations, the LEP
                  Electroweak Working Group], hep-ex/0511027;\\
                  F.~Spano, 
                  talk given at {\em Rencontres de Moriond,
                Electroweak Interactions and Unified Theories}, 
                   La Thuile, Italy, March 2006;\\
                   see also: {\tt lepewwg.web.cern.ch/LEPEWWG/Welcome.html}.

\bibitem{LEPHiggsSM} [LEP Higgs working group], 
                     {\em Phys. Lett.} {\bf B 565} (2003) 61, 
                     hep-ex/0306033.

\bibitem{mhiggslong} S.~Heinemeyer, W.~Hollik and G.~Weiglein,
                     {\em Eur. Phys. J.} {\bf C 9} (1999) 343,
                     hep-ph/9812472.

\bibitem{mhiggsAEC} G.~Degrassi, S.~Heinemeyer, W.~Hollik,
                    P.~Slavich and G.~Weiglein, 
                    {\em Eur. Phys. J.} {\bf C 28} (2003) 133,
                    hep-ph/0212020.


\bibitem{Allanach:2004rh}
                   B.~Allanach, A.~Djouadi, J.~Kneur, W.~Porod and P.~Slavich,
                   {\em JHEP} {\bf 0409} (2004) 044, 
                   hep-ph/0406166.

\bibitem{diffHSM} M.~Albrow and A.~Rostovtsev,
                  hep-ph/0009336;\\
                  V.~Khoze, A.~Martin and M.~Ryskin,
                  {\em Eur. Phys. J.} {\bf C 23} (2002) 311, 
                  hep-ph/0111078;\\
                  A.~De~Roeck, V.~Khoze, A.~Martin, R.~Orava and M.~Ryskin, 
                  {\em Eur. Phys. J.} {\bf C 25} (2002) 391,
                  hep-ph/0207042;\\
                  B.~Cox,
                  {\em AIP Conf.\ Proc.} {\bf 753} (2005) 103, 
                  hep-ph/0409144;\\
                  J.~Forshaw,
                  hep-ph/0508274.

\bibitem{diffHMSSM} S.~Heinemeyer, V.~Khoze, M.~Ryskin, W.~Stirling,
                    M.~Tasevsky and G.~Weiglein,
                    {\em in preparation}. 

\bibitem{Carena:2000yx}
M.~Carena et al.  [Higgs Working Group Collaboration],
  hep-ph/0010338.



\bibitem{Harlander:2002wh}
  R.~Harlander and W.~Kilgore,
  {\em Phys.\ Rev.\ Lett.} {\bf 88} (2002) 201801,
  hep-ph/0201206.

\bibitem{Ravindran:2003um}
  V.~Ravindran, J.~Smith and W.~van Neerven,
  {\em Nucl.\ Phys.} {\bf B 665} (2003) 325, 
  hep-ph/0302135.

\bibitem{Anastasiou:2002yz}
  C.~Anastasiou and K.~Melnikov,
  {\em Nucl.\ Phys.} {\bf B 646} (2002) 220,
  hep-ph/0207004.

\bibitem{Graudenz:1992pv}
  D.~Graudenz, M.~Spira and P.~Zerwas,
  {\em Phys.\ Rev.\ Lett.} {\bf 70} (1993) 1372;\\
  M.~Spira, A.~Djouadi, D.~Graudenz and P.~Zerwas,
  {\em Nucl.\ Phys.} {\bf B 453} (1995) 17,
  hep-ph/9504378.

\bibitem{Moch:2005ky}
  S.~Moch and A.~Vogt,
  {\em Phys.\ Lett.} {\bf B 631} (2005) 48, 
  hep-ph/0508265.

\bibitem{Laenen:2005uz}
  E.~Laenen and L.~Magnea,
  {\em Phys.\ Lett.} {\bf B 632} (2006) 270, 
  hep-ph/0508284.

\bibitem{Catani:2003zt}
  S.~Catani, D.~de Florian, M.~Grazzini and P.~Nason,
  {\em JHEP} {\bf 0307} (2003) 028,
  hep-ph/0306211.

\bibitem{Martin:2002aw}
  A.~Martin, R.~Roberts, W.~Stirling and R.~Thorne,
  {\em Eur.\ Phys.\ J.} {\bf C 28} (2003) 455,
  hep-ph/0211080.

\bibitem{Anastasiou:2004xq}
  C.~Anastasiou, K.~Melnikov and F.~Petriello,
  {\em Phys.\ Rev.\ Lett.} {\bf 93} (2004) 262002,
  hep-ph/0409088.

\bibitem{Aglietti:2004nj}
  U.~Aglietti, R.~Bonciani, G.~Degrassi and A.~Vicini,
  {\em Phys.\ Lett.} {\bf B 595} (2004) 432,
  hep-ph/0404071.

\bibitem{Degrassi:2004mx}
  G.~Degrassi and F.~Maltoni,
  {\em Phys.\ Lett.} {\bf B 600} (2004) 255,
  hep-ph/0407249.


\bibitem{Han:1992hr}
  T.~Han, G.~Valencia and S.~Willenbrock,
  {\em Phys.\ Rev.\ Lett.} {\bf 69} (1992) 3274,
  hep-ph/9206246.

\bibitem{Berger:2004pc}
  E.~Berger and J.~Campbell,
  {\em Phys.\ Rev.} {\bf D 70} (2004) 073011,
  hep-ph/0403194.

\bibitem{Figy:2003nv}
  T.~Figy, C.~Oleari and D.~Zeppenfeld,
  {\em Phys.\ Rev.} {\bf D 68} (2003) 073005,
  hep-ph/0306109.

\bibitem{Campbell:1999ah}
  J.~Campbell and R.~Ellis,
  {\em Phys.\ Rev.} {\bf D 60} (1999) 113006,
  hep-ph/9905386.

\bibitem{Pumplin:2002vw}
  J.~Pumplin, D.~Stump, J.~Huston, H.~Lai, P.~Nadolsky and W.~Tung,
  {\em JHEP} {\bf 0207} (2002) 012,
  hep-ph/0201195.

\bibitem{Brein:2003wg}
  O.~Brein, A.~Djouadi and R.~Harlander,
  {\em Phys.\ Lett.} {\bf B 579} (2004) 149, 
  hep-ph/0307206.


\bibitem{Ciccolini:2003jy}
  M.~Ciccolini, S.~Dittmaier and M.~Kramer,
  {\em Phys.\ Rev.} {\bf D 68} (2003) 073003,
  hep-ph/0306234.

\bibitem{Assamagan:2004mu}
  K.~Assamagan et al.  [Higgs Working Group Collaboration],
  hep-ph/0406152.



\bibitem{Harlander:2003ai}
  R.~Harlander and W.~Kilgore,
  {\em Phys.\ Rev.} {\bf D 68} (2003) 013001,
  hep-ph/0304035.

\bibitem{Dittmaier:2003ej}
  S.~Dittmaier, M.~Kramer and M.~Spira,
  {\em Phys.\ Rev.} {\bf D 70} (2004) 074010, 
  hep-ph/0309204.


\bibitem{Dawson:2003kb}
  S.~Dawson, C.~Jackson, L.~Reina and D.~Wackeroth,
  {\em Phys.\ Rev.} {\bf D 69} (2004) 074027, 
  hep-ph/0311067.

\bibitem{Dawson:2004sh}
  S.~Dawson, C.~Jackson, L.~Reina and D.~Wackeroth,
  {\em Phys.\ Rev.\ Lett.} {\bf 94} (2005) 031802, 
  hep-ph/0408077.

\bibitem{Dawson:2005vi}
  S.~Dawson, C.~Jackson, L.~Reina and D.~Wackeroth,
  {\em Mod.\ Phys.\ Lett.} {\bf A 21} (2006) 89, 
  hep-ph/0508293.

\bibitem{Campbell:2002zm}
  J.~Campbell, R.~Ellis, F.~Maltoni and S.~Willenbrock,
  {\em Phys.\ Rev.} {\bf D 67} (2003) 095002,
  hep-ph/0204093.


\bibitem{Beenakker:2001rj}
  W.~Beenakker, S.~Dittmaier, M.~Kramer, B.~Plumper, M.~Spira and P.~Zerwas,
  {\em Phys.\ Rev.\ Lett.} {\bf 87} (2001) 201805, 
  hep-ph/0107081.

\bibitem{Reina:2001sf}
  L.~Reina and S.~Dawson,
  {\em Phys.\ Rev.\ Lett.} {\bf 87} (2001) 201804,
  hep-ph/0107101.

\bibitem{Dawson:2002tg}
  S.~Dawson, L.~Orr, L.~Reina and D.~Wackeroth,
  {\em Phys.\ Rev.} {\bf 67} (2003) 071503, 
  hep-ph/0211438.

\bibitem{Ballestrero:1992bk}
  A.~Ballestrero and E.~Maina,
  {\em Phys.\ Lett.} {\bf B 299} (1999) 312. 

\bibitem{Maltoni:2001hu}
  F.~Maltoni, K.~Paul, T.~Stelzer and S.~Willenbrock,
  {\em Phys.\ Rev.} {\bf D 64} (2001) 094023, 
  hep-ph/0106293.


\bibitem{LEPHiggsMSSM} [LEP Higgs working group],
                   hep-ex/0602042.

\bibitem{D0bounds} V.~Abazov et al.  [D0 Collaboration],
                   hep-ex/0504018.

\bibitem{CDFbounds} A.~Abulencia et al.  [CDF Collaboration],
                    hep-ex/0508051.

\bibitem{Tevcharged} [CDF Collaboration],
                     hep-ex/0510065;\\
                     R.~Eusebi, Ph.d. thesis: ``Search for charged Higgs 
                     in $t \bar t$ decay products from proton-antiproton
                     collisions at $\sqrt{s}=1.96\,{\rm TeV}$'', 
                     University of Rochester, 2005.

\bibitem{schumi} M.~Schumacher,
                 {\em Czech. J. Phys.} {\bf 54} (2004) A103;
                 hep-ph/0410112.

\bibitem{benchmark} M.~Carena, S.~Heinemeyer, C.~Wagner and G.~Weiglein,
                    hep-ph/9912223.

\bibitem{benchmark2} M.~Carena, S.~Heinemeyer, C.~Wagner and G.~Weiglein, 
                     {\em Eur. Phys. J.} {\bf C 26} (2003) 601, 
                     hep-ph/0202167.

\bibitem{benchmark3} M.~Carena, S.~Heinemeyer, C.~Wagner and G.~Weiglein,
                     {\em Eur.\ Phys.\ J.} {\bf C 45} (2006) 797, 
                     hep-ph/0511023.

\bibitem{feynhiggs} S.~Heinemeyer, W.~Hollik and G.~Weiglein,
                    {\em Comput. Phys. Comm.} {\bf 124} (2000) 76,
                    hep-ph/9812320;
                    hep-ph/0002213;\\
                    M.~Frank, S.~Heinemeyer, W.~Hollik and 
                    G.~Weiglein,
                    hep-ph/0202166;\\
                    T.~Hahn, S.~Heinemeyer, W.~Hollik and G.~Weiglein,
                    hep-ph/0507009;\\
                    M.~Frank, T.~Hahn, S.~Heinemeyer, W.~Hollik,  
                    H.~Rzehak and G.~Weiglein,
                    {\em in preparation};\\
                    see: {\tt www.feynhiggs.de} .

\bibitem{mhiggsletter} S.~Heinemeyer, W.~Hollik and G.~Weiglein,
                     {\em Phys. Rev.} {\bf D 58} (1998) 091701, 
                     hep-ph/9803277; 
                     {\em Phys. Lett.} {\bf B 440} (1998) 296, 
                     hep-ph/9807423.

\bibitem{bse} M.~Carena, H.~Haber, S.~Heinemeyer, W.~Hollik, C.~Wagner,
              and G.~Weiglein,
              {\em Nucl. Phys.} {\bf B 580} (2000) 29,
              hep-ph/0001002.


\bibitem{mt1727} [The Tevatron Electroweak Working Group],
                 hep-ex/0507091.

\bibitem{tbexcl} S.~Heinemeyer, W.~Hollik and G.~Weiglein, 
                 {\em JHEP} {\bf 0006} (2000) 009,
                 hep-ph/9909540.

\bibitem{ggsuppr} A.~Djouadi,
                  {\em Phys. Lett.} {\bf B 435} (1998) 101,
                  hep-ph/9806315.



\bibitem{gghMSSM2L} R.~Harlander and M.~Steinhauser,
                    {\em JHEP} {\bf 0409} (2004) 066, 
                    hep-ph/0409010.

\bibitem{Hahn:2002gm} T.~Hahn, S.~Heinemeyer and G.~Weiglein,
                      {\em Nucl.\ Phys.} {\bf B 652} (2003) 229,
                      hep-ph/0211204.





\end{thebibliography}


\end{document}